\documentstyle[12pt,seceqn,epsf]{article}

\begin{document}

\title{ Boundary conformal field theory approach to
the critical two-dimensional Ising model with a defect line}

\author{Masaki
Oshikawa$^{a,}$\thanks{E-mail: {\tt oshikawa@physics.ubc.ca}}
and
Ian Affleck$^{a,b,}$\thanks{E-mail: {\tt affleck@physics.ubc.ca}} \\
\\
\it
\protect{$^a$}Department of Physics and Astronomy,
University of British Columbia \\
\it  Vancouver BC V6T 1Z1 Canada
\\
\it \protect{$^b$}Canadian Institute for Advanced Research,
University of British Columbia \\
\it Vancouver BC V6T 1Z1 Canada }

\date{Dec 18,1996}

\maketitle

\begin{abstract}
We study the critical two-dimensional Ising model with a defect line
(altered bond strength along a line) in the continuum limit.
By folding the system at the defect line, the problem is mapped
to a special case of the critical Ashkin-Teller model, 
the continuum limit of which is the $Z_2$ orbifold of the free boson,
with a boundary.
Possible boundary states on the $Z_2$ orbifold theory
are explored, and a special case is applied to the Ising defect problem.
We find the complete spectrum of boundary operators,
exact two-point correlation functions
and the universal term in the free energy of the defect line
for arbitrary strength of the defect.
We also find a new universality class of defect lines.
It is conjectured that we have found all the possible universality
classes of defect lines in the Ising model.
Relative stabilities among the defect universality classes are
discussed.
\end{abstract}

\bigskip \bigskip

\section{Introduction}

Boundary (surface) critical phenomena~\cite{Binder:surface} in
two-dimensional systems have attracted a lot of interest.
Even if the bulk system is a free system,
the presence of the boundary makes the problem non-trivial.
Boundary conformal field theory~\cite{Cardy:BCFT}
is a powerful tool to
solve these problems. Its applications to quantum
impurity problems are also discussed~\cite{Affleck:reviewKondo}.
A variation of the problem
has a defect line {\em inside} a two-dimensional system.
This kind of problem can still be mapped to a system with boundary,
by folding the system at the defect line~\cite{WongAffleck}.

The two-dimensional Ising model, which was solved 50 years ago by
Onsager~\cite{Onsager:Ising},
is still useful to test new techniques and ideas in
statistical mechanics.
In this paper, we apply boundary conformal field theory
to the two-dimensional Ising model with a defect line.
This problem has been studied by several authors.
Bariev~\cite{Bariev:Ising}
found a continuously varying ``surface'' critical exponent.
McCoy and Perk~\cite{McCoyPerk} and Kadanoff~\cite{Kadanoff:defect}
discussed the correlation functions along the defect line.
Brown~\cite{Brown:defect} discussed several properties of this model.
In particular, he calculated two-point spin correlation function
for general locations, in first-order perturbation of the
defect strength.
Several other studies are focused on 
the finite-size scaling of the transfer-matrix spectrum
on a cylinder.
Cabrera and Julien~\cite{CabreraJulien} calculated the spectrum
numerically.
Following the application of conformal invariance by
Turban~\cite{Turban:defect} (see also Ref.~\cite{Drugowich}),
Henkel~{\it et. al.}~\cite{Henkel87,Henkel89} studied the spectrum
of the quantum version by various methods and discussed the
spectrum in terms of Virasoro and Kac-Moody algebras.
Exact formulae for the spectrum on the square lattice with general
anisotropy are given by
Abraham~{\it et. al.}~\cite{AbrahamKoSvrakic}.
Grimm~\cite{Grimm} also examined generalizations of the defect line.
Recently,
an $S$-matrix approach to the problem was also presented by
Delfino~{\em et. al.}~\cite{Delfino:defect}.  

In this paper, 
we study the continuum limit at the critical point.
To relate this problem to boundary conformal field theory,
we ``fold'' the system into a $c=1$ conformal field theory with a boundary.
Using the exact partition function, we identify the boundary states
in terms of $c=1$ conformal field theory.
Folding the Ising model gives two decoupled Ising models in the
bulk, which is a special case of the Ashkin-Teller model.
The critical theory of the Ashkin-Teller model is a $c=1$ conformal
field theory: the $Z_2$ orbifold of a free boson.
We explore the possible boundary states of the orbifold theory
and apply a special case to the Ising defect problem.

Identification of the boundary states
enables us to give the complete spectrum of ``surface''
critical exponents for general strength of the defect.
Cardy and Lewellen~\cite{CardyLewellen} showed
that the boundary correlation functions can be determined
from the boundary state and the Operator Product
Expansion (OPE) in the bulk.
We employ this idea to calculate the exact two-point spin
correlation function,
using Zamolodchikov's solution~\cite{Zam:ATspin} for $c=1$ conformal
blocks of spin operators.
We also calculate certain four-point correlation functions,
as well as some correlation
function near the end of the defect line.
A universal term in the free energy with a finite length defect
is also given.
Moreover, we find a new one-parameter universality class of defect
lines in terms of boundary conformal field theory.
We calculate two-point spin correlation functions and then identify
a corresponding defect in the quantum lattice model.
Finally, the relative stability of the defect universality classes
is discussed, and a conjecture is made regarding the complete set of
defect universality classes.

The organization of the paper is as follows.
In Section~\ref{sec:partitionfn},
we derive the partition function of the model
in the critical continuum limit.
We use the result to identify the boundary condition in terms
of Ashkin-Teller model in Section~\ref{sec:bstate}.
The possible boundary states on the $Z_2$ orbifold theory is
explored in Section~\ref{sec:orbstate}, and further applied to
the present problem.
In Section~\ref{sec:Dcorr}, we calculate
two-point correlation functions and other universal quantities
for arbitrary strength of the defect.
A new one-parameter family of defect lines is studied
in Section~\ref{sec:Ncorr}.
In Section~\ref{sec:RGflow}, we discuss the relative stability and
renormalization-group flow among the defect universality classes.
We give a summary and some discussion in Section~\ref{sec:summary}.

A brief description of the present work has appeared
in~\cite{IsingLetter}.

\section{Partition function at the critical point}
\label{sec:partitionfn}

Here we derive the partition function of the Ising model with a
defect line, in the continuum limit.
While Henkel~{\it et. al.}~\cite{Henkel87,Henkel89}
have found the result in the $\tau$-continuum limit
(quantum Ising chain), here we derive the partition
function for the square lattice Ising model with an arbitrary
anisotropy, taking the continuum limit of the analysis
by Abraham {\it et. al.}~\cite{AbrahamKoSvrakic}.
Apart from the trivial non-universal parts,
the result is independent of the anisotropy, under an appropriate
rescaling. It agrees with that of Henkel {\it et. al.}
which corresponds to the extremely anisotropic limit.
This is a manifestation of the universality in the Ising model
in the presence of a defect line.
 
We consider a square-lattice Ising model with the lattice constant $a$.
The Ising model on a cylinder with a defect line is defined by
the classical Hamiltonian:
\begin{equation}
\label{eq:defectham}
  {\cal E} = - \sum_{i=1,M-1} \sum_{j=1,N}
           [   J_1 \sigma_{i,j} \sigma_{i,j+1}
              + J_2 \sigma_{i,j} \sigma_{i+1,j} ]
        - \sum_{j=1,N}
           [   J_1 \sigma_{M,j} \sigma_{M,j+1}
              + \tilde{J} \sigma_{M,j} \sigma_{1,j} ] .
\end{equation}
The altered link (defect) is placed between $i=M$ and $i=1$.
We denote the defect strength by $b = \tilde{J} / J_2$.
This model 
reduces to the periodic boundary condition, the free boundary
condition and the antiperiodic boundary condition,
respectively for $b=1,0,-1$.
We also introduce $K_1, K_2$ and $K^*_1$ by
\begin{equation}
  K_1 = J_1 / kT , K_2 = J_1 / k T
\end{equation}
\begin{equation}
\sinh{K^*_1} \sinh{K_1} = 1 .  
\end{equation}

Let us refer to the direction parallel to the defect line as horizontal.
For periodic or antiperiodic boundary condition the transfer
matrix can be mapped to a free-fermion Hamiltonian~\cite{SchultzMattisLieb}
by the Jordan-Wigner transformation.
However, it is not straightforward for general $b$ because
the system lacks translation invariance in the
$i$ (vertical) direction
and a simple Fourier transformation is not useful.
Nevertheless, Abraham, Ko and Svrakic~\cite{AbrahamKoSvrakic}
obtained the spectrum of the transfer matrix
in the $j$ (horizontal) direction using a spinor approach.
We start from their result and apply it to the critical case.

According to their result,
the transfer matrix $V$ can be simply expressed by fermion operators,
but the Hilbert space is divided into two sectors:
\begin{equation}
  V = P_+ V_+ + P_- V_- .
\label{eq:sectors}
\end{equation}
Here $P_+$ ($P_-$) is the projection operator onto the even (odd)
fermion-number sector. However, one must be careful about the definition
of the ``even'' and ``odd'' sectors. (See below.)
The transfer matrices in the two sectors are given by
\begin{equation}
V_{\pm} = \exp{[ - \sum_{j=1}^{M}
    \gamma( a k^{\pm}_j) ( c^{\dagger}_j c_j -1/2) ]}.  
\end{equation}
The ``single-fermion energy'' $\gamma$ is given by the
Onsager dispersion function
\begin{equation}
  \cosh{\gamma( p )} \equiv
   \cosh{2 K^*_1} \cosh{2 K_2} - \sinh{2 K^*_1} \sinh{2 K_2} \cos{p}
\end{equation}
and the quantization condition of the wave number $0 < k < \pi / a$
\begin{eqnarray}
  e^{i M a k^{\pm}} &=&  \zeta e^{ i \alpha (a k^{\pm} , \mp \zeta b)} \\
  \tan{\frac{\alpha(p,x)}{2}} &\equiv&
      \frac{\sinh{[(1-x)K_2]}}{\sinh{[(1+x)K_2]}}
            \tan{\frac{\delta^*(p)}{2}}
\label{eq:qcond}
\end{eqnarray}
where $\pm$ in the first equation corresponds to $V_{\pm}$,
$a$ is the lattice constant,
$\zeta = + 1$ or $-1$ which is independent of $V_{\pm}$.
$\delta^*$ is defined as
\begin{equation}
  \sinh{\gamma(p)} \cos{\delta^*(p)} =
     \sinh{2 K_2} \cosh{2 K_1^*} -
     \cosh{2 K_2} \sinh{2 K_1^*} \cos{p} .
\end{equation}

Let us discuss the critical point where $K_2 = K^*_1$
but $b$ still remains as a free parameter.
In order to discuss the universal behavior at large distances,
we take the continuum limit $a \rightarrow 0$
(or equivalently consider the scaling limit where the length scale
is much larger than the lattice constant),
keeping the circumference of the cylinder $L = Ma$ constant.

Then only the low-energy limit of the dispersion relation
is relevant in the discussion.
It is given by
\begin{equation}
  \gamma(k) = v | k |,
\label{eq:lindisp}
\end{equation}
where $v$ is a constant (spin wave velocity).
In the following, we rescale the horizontal (``imaginary time'')
direction so that $v=1$.
The transfer matrix can be described by a Hamiltonian in the
continuum limit.
The Hamiltonians for fermion number even/odd sector are given by
\begin{equation}
H_{\pm} =  \sum_{j}
            | k^{\pm}_j | ( c^{\dagger}_j c_j -1/2) .  
\label{eq:isingham}
\end{equation}

In this limit, $\delta^*(k) = \pi/2$ independent of $k$ and
the quantization condition~(\ref{eq:qcond}) is much simplified.
Furthermore, we can choose freely the sign of $k$ in the discussion
of the energy,
because the dispersion function is an even function of $k$.
Thus the quantization condition in the continuum limit is written as
\begin{equation}
  k^{\pm} = \frac{\alpha(\mp b)}{2\pi} + \frac{2 \pi}{L} n
\label{eq:phshift}
\end{equation}
where $n$ is an arbitrary integer and $\alpha$ is now independent
of $k$:
\begin{equation}
  \tan{\frac{\alpha(x)}{2}} =
      \frac{\sinh{[(1-x)K_2]}}{\sinh{[(1+x)K_2]}} .
\label{eq:alphadef}
\end{equation}
From this equation, we see that $\alpha(x)$ satisfies
\begin{equation}
  \alpha(x) + \alpha(-x) = \pi .
\end{equation}
$\alpha(x)$ takes the values $\pi$, $\pi/2$ and $0$ respectively for
$x=-1$,$0$ and $1$.
When $x \rightarrow \pm \infty$,
$\alpha$ is given by
\begin{equation}
  \tan{\frac{\alpha(\pm \infty)}{2}} = - e^{ \mp 2 K_2} .
\end{equation}
Thus $\pi < \alpha(-\infty) < 3 \pi /2$ and $ - \pi/2 < \alpha(\infty) < 0$.
We note that, in the $K_2 \rightarrow 0$ limit
with the $x \rightarrow \pm \infty$ limit, $\alpha$ approaches
the limiting values $\alpha(-\infty) = 3 \pi /2$ and
$\alpha(\infty) = -\pi/2$.
The $K_2 \rightarrow 0$ limit
corresponds to the anisotropic limit where the horizontal link becomes
very strong. (It should be remembered that we stay at the critical point
where $K_2 = K_1^*$.)
Regarding the horizontal direction as imaginary time, this
is the so-called $\tau$-continuum limit~\cite{FradkinSusskind}
of the Ising model.
In this limit, the system is equivalent to the one-dimensional
quantum Ising model in a transverse field.
The quantum Hamiltonian is given by
\begin{equation}
\label{eq:qham}
  H = - \sum_{n = - \infty}^{\infty} \hat{\sigma}^x(n)
      - \sum_{n \neq 0} \hat{\sigma}^z(n-1) \hat{\sigma}^z(n)
      - b \hat{\sigma}^z(-1) \hat{\sigma}^z(0), 
\end{equation}
where $\hat{\sigma}^{x,z}(n)$ is a Pauli spin operator at site $n$.
This model is the critical transverse Ising model with a defective
link between $n=-1$ and $0$.

In eq.~(\ref{eq:isingham}), we arrived at a rather simple description
of the transfer matrix
for generalized boundary condition with parameter $b$.
Namely, it is always a free fermion Hamiltonian
and the dispersion relation~(\ref{eq:lindisp}) is independent of $b$.
The only effect of the boundary condition is the ``phase shift''
$\alpha$ of the wavenumber as in~(\ref{eq:phshift}).
This phase shift is different in even- and odd- fermion number
sectors.

We note that the parity of the fermion number is a subtle problem.
Exchanging a fermion annihilation operator $c_k$ and the corresponding
creation operator $c^{\dagger}_k$ preserves all the fermion anticommutation
relations. However, this procedure flips the fermion number parity operator
defined by $(-1)^F = (-1)^{\sum_j c^{\dagger}_j c_j}$.
Thus the fermion parity operator depends on the definition of
the fermion creation/annihilation operators.
We fix the fermion operators so that all particles have non-negative
energy and the Hamiltonian takes the form of eq.~(\ref{eq:isingham}).
For $\alpha >0$, the two sectors of Hilbert space in~(\ref{eq:sectors})
actually corresponds to the even/odd fermion number in the above definition:
\begin{equation}
  P_+ = \frac{ 1 + (-1)^F}{2} , \hfill
  P_- = \frac{ 1 - (-1)^F}{2} .
\label{eq:parityop}
\end{equation}
When $\alpha$ changes sign,
one of the quantized wavevectors passes through $k =0$.
The energy of the corresponding fermion seems to change non-analytically
as $|k|$.
However, this should be understood as the fermion energy depending on
on the parameter linearly but the annihilation/creation operators
being exchanged when the single fermion energy becomes negative.
Hence the fermion number parity should be flipped when
$\alpha$ becomes negative (as long as we keep the above definition
of the fermion operators.)

\begin{figure}[htbp]
  \begin{center}
    \leavevmode
    \epsfxsize=\textwidth
    \epsfbox{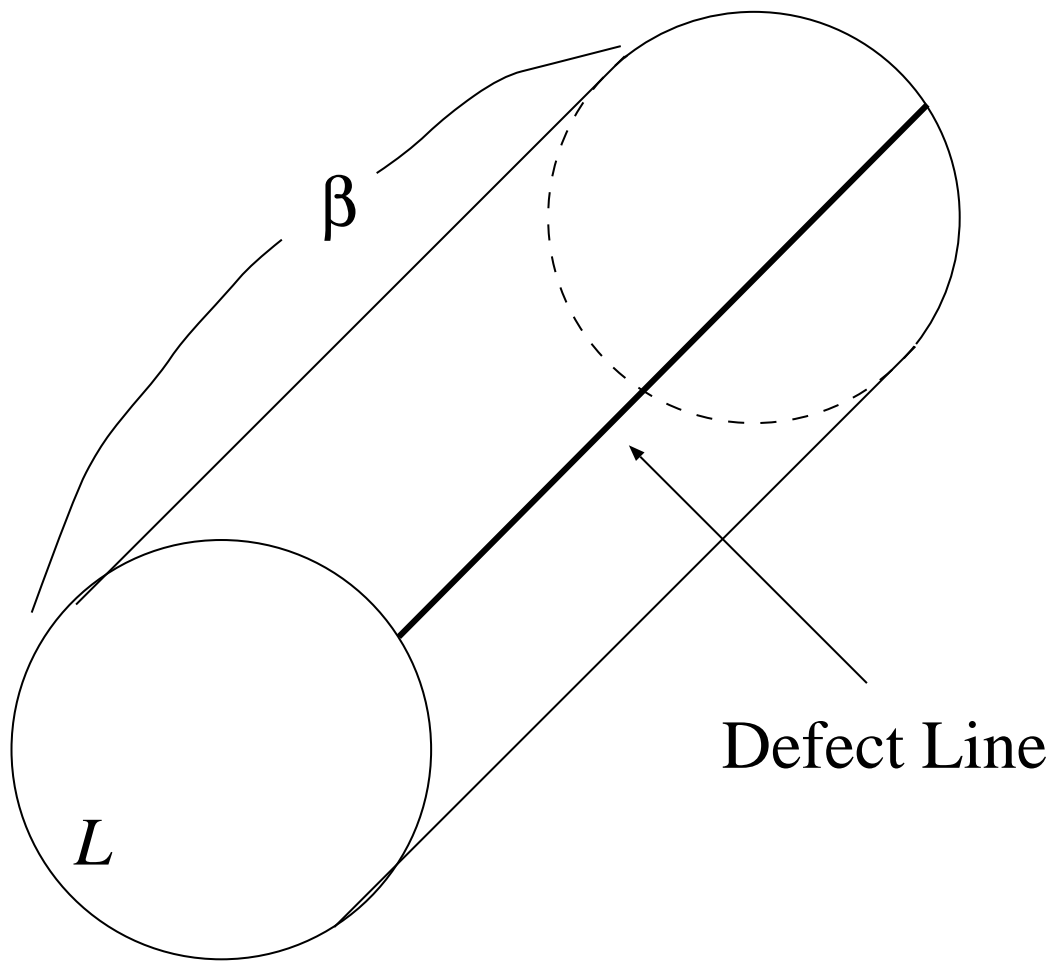}
    \caption{The Ising model the defect line on a cylinder.
The defect line is parallel to the cylinder axis.}            
    \label{fig:Isingfigure}
  \end{center}
\end{figure}

Let us consider the partition function of the Ising model on a
cylinder with circumference $L$ and length $\beta$ (a macroscopic length
scale), as shown in Fig.~\ref{fig:Isingfigure}.
We apply the periodic boundary condition in the $\beta$ direction.
The partition function is given by
\begin{equation}
  Z = {\rm Tr} \exp{(- \beta H)}
\end{equation}
where $H = P_+ H_+ + P_- H_-$ is the total Hamiltonian.

First we consider the $-1 \leq b \leq 1$ case.
Here $\alpha(\pm b) \geq 0$ and thus the assignment~(\ref{eq:parityop})
is valid.
In the following we denote $\alpha(b)$ by simply $\alpha$:
$0 \leq \alpha \leq \pi$ for $ -1 \leq b \leq 1$.
In the odd ($(-1)^F = - 1$) sector, the fermion one-particle
energies are given by 
$\epsilon_n = |( \alpha + 2\pi n )/L |$.
The ground state is the fermionic vacuum $| \mbox{vac} \rangle$
which satisfies $c_k | \mbox{vac} \rangle = 0$.
The vacuum still has ``zero-point energy''
\begin{equation}
  E_{\rm vac} = - \frac{1}{2}  \sum_n \epsilon_n
\end{equation}
according to eq.~(\ref{eq:isingham}).
While this quantity apparently diverges (but of course is cut off by the
lattice) in the continuum limit, we can extract the universal
``Casimir energy'' in the odd sector as
\begin{equation}
  E^-_{\rm vac} = C L + A +
     \frac{1}{L}
     \left( \frac{\pi}{6} + \frac{\alpha^2}{4 \pi} - \frac{\alpha}{2} \right)
\label{eq:oddvac}
\end{equation}
where $C$ is a non-universal energy density, which is
the order of inverse square of the lattice constant. 
The second term is a non-universal shift of the ground-state energy
due to the defect. (``surface energy'').
The last term is the universal Casimir energy.

For even sector, we can simply replace $\alpha$ by $ \pi - \alpha$
to obtain
\begin{equation}
  E^+_{\rm vac} = C L + A +
     \frac{1}{L}
     \left( - \frac{\pi}{12} + \frac{\alpha^2}{4 \pi} \right) .
\label{eq:evenvac}
\end{equation}
Since there must be at least one fermion in the odd sector,
the ground state energy in the odd sector is given by
\begin{equation}
  E^-_g = E^-_{\rm vac} + \frac{ \pi - \alpha}{L}
\end{equation}
The ground state energy in the even sector is simply~(\ref{eq:evenvac}) 
and is lower than $E^-_g$.
Thus eq.~(\ref{eq:evenvac}) also gives the ground-state
energy of the present system.
It reduces to the known value for periodic ($\alpha =0$),
antiperiodic ($\alpha = \pi$) and free ($\alpha = \pi/2$)
cases~\cite{Cardy:bceffect}.

In the following we ignore the non-universal part and
set $L=1$ for simplicity. (The $L$ dependence
can be recovered by simply replacing $\beta \rightarrow \beta/L$.)
The partition function in the odd sector is given by
\begin{equation}
  Z_{\rm odd} = e^{- \beta E_{\rm vac}}
    \sum_{\{n_k = 0,1\}} P_- e^{ - \beta \sum_k \epsilon_k n_k } 
\end{equation}
Using
\begin{equation}
  P_- = \frac{ 1 - (-1)^F}{2} = \frac{ 1 - (-1)^{\sum_k n_k}}{2},
\end{equation}
we obtain the infinite product representation of the elliptic theta
functions as
\begin{equation}
  Z_{\rm odd} = \frac{e^{-\alpha^2 \beta/4\pi}}{2}
                \left[ \frac{\vartheta_2(w,q)}{\eta(q)} 
                 + i \frac{\vartheta_1(w,q)}{\eta(q)} \right] ,
\label{eq:oddZ}
\end{equation}
where the parameters is given by
\begin{equation}
q = e^{-2 \pi \beta}, \hfill w = e^{- \alpha \beta}.  
\label{eq:qandw}
\end{equation}
In this paper we use the elliptic theta functions
and Dedekind eta function
\begin{eqnarray}
\vartheta_1(w,q) &=& i \sum_{n = - \infty}^{\infty}
              (-1)^n q^{\frac{1}{2}(n-\frac{1}{2})^2} w^{n-\frac{1}{2}} \\
\vartheta_2(w,q) &=&  \sum_{n = - \infty}^{\infty}
                 q^{\frac{1}{2}(n-\frac{1}{2})^2} w^{n-\frac{1}{2}} \\
\vartheta_3(w,q) &=&  \sum_{n = - \infty}^{\infty}
                 q^{\frac{1}{2}n^2} w^{n} \\
\vartheta_4(w,q) &=&  \sum_{n = - \infty}^{\infty}
                       (-1)^n q^{\frac{1}{2}n^2} w^{n} \\
\eta(q) &=&   q^{1/24} \prod_{n=1}^{\infty} (1 - q^{n}).
\label{eq:theta}
\end{eqnarray}
Theta functions with a single argument $\vartheta_j(q)$ should be
understood as $\vartheta_j(w=1,q)$ (``theta constants'').

For the even sector we make a similar calculation and obtain
\begin{equation}
  Z_{\rm even} = \frac{e^{-\alpha^2 \beta/4\pi}}{2}
                \left[ \frac{\vartheta_3(w,q)}{\eta(q)} 
                 + \frac{\vartheta_4(w,q)}{\eta(q)} \right] .
\label{eq:evenZ}
\end{equation}
Thus the total partition function $Z = Z_{\rm even} + Z_{\rm odd}$
is given by
\begin{equation}
  Z_{\rm Ising} = \frac{e^{-\alpha^2 \beta/4\pi}}{2}
                \left[ \frac{\vartheta_3(w,q)}{\eta(q)} 
                 + \frac{\vartheta_4(w,q)}{\eta(q)}
                 + \frac{\vartheta_2(w,q)}{\eta(q)} 
                 + i \frac{\vartheta_1(w,q)}{\eta(q)} \right] ,
\label{eq:totalZ}
\end{equation}
with the parameters as in eq.~(\ref{eq:qandw}).

For $|b| >1$, we must flip the fermion parity in the sector with
negative phase shift $\alpha$.
However, as a result of the calculation,
we found that the above equation~(\ref{eq:totalZ}) is
still valid in this case.
Thus the partition function of the critical Ising model on the cylinder
is given by eq.~(\ref{eq:totalZ}) for
the entire range $-\infty < b < \infty$.
Our result agrees with that by Henkel {\it et. al.}~\cite{Henkel87,Henkel89},
which is obtained in $\tau$-continuum limit.

\section{Identification of the boundary state by folding}
\label{sec:bstate}

\begin{figure}[htbp]
  \begin{center}
    \leavevmode
    \epsfysize=12cm
    \epsfbox{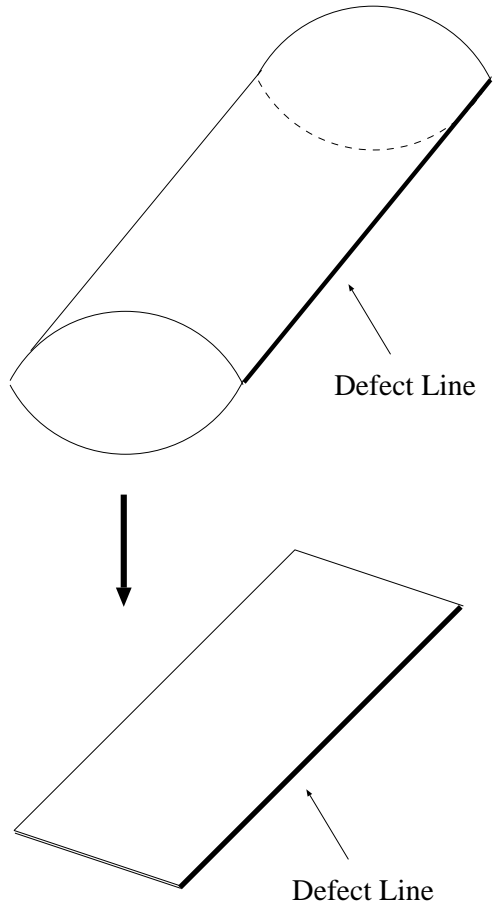}
    \caption{The folding of the Ising model on a cylinder to
a $c=1$ theory on a strip. We fold at the defect line and
also at the line on the opposite side.
These lines correspond to the boundary in the folded system.}
    \label{fig:fold}
  \end{center}
\end{figure}

In order to apply
boundary conformal field theory to the present problem,
we fold the system so that the defect line becomes the boundary
of the system~\cite{WongAffleck}.
The cylinder of circumference $1$ is folded to a strip of
width $1/2$, as shown in Fig.~\ref{fig:fold}.
In addition to the boundary which corresponds to the defect line,
we have another boundary on the opposite side.
The two boundaries are identical when there is no defect line, or
equivalently when the periodic boundary condition is imposed
on the Ising model ($b=1$).
As a result of the folding, the degrees of freedom in the bulk
are doubled. There are two different kinds of spin operators at each point,
corresponding to the locations before the folding.
We will call them two different layers of spins.
Because of the doubling, we must consider a $c=1$ conformal field theory,
rather than $c=1/2$.
Known $c=1$ conformal field theories are described by a free boson
or its orbifold.

Actually, a useful trick
to study the Ising model on a plane is to regard two
independent Ising models as
a free boson theory~\cite{LutherPeschel,ZuberItzykson:Ising}.
Let us denote the spins in each layer as $\sigma_1$ and $\sigma_2$.
The product $\sigma_1 \sigma_2$
of the spin operators in each layer (at the same point)
can be expressed as a simple bosonic operator $\cos{\varphi}$ where
$\varphi$ is the free boson field.
Since the two Ising models are independent, correlation functions
of this composite operator is always the square of the spin
correlation in a single layer.
Thus we can calculate spin correlation functions of the Ising model
in terms of the free boson.
In the present problem, the two Ising layers are coupled at the boundary
though they are independent in the bulk.
This makes the problem more difficult.

The two-dimensional Ashkin-Teller model~\cite{AshkinTeller}
is defined by the classical Hamiltonian
\begin{equation}
  {\cal E} = \sum_{\langle x y \rangle}
 \left\{
       K[ \sigma_1(x) \sigma_1(y) + \sigma_2(x) \sigma_2(y) ]
    +  L \sigma_1(x) \sigma_1(y) \sigma_2(x) \sigma_2(y)
 \right\} ,
\end{equation}
where $\langle x, y \rangle$ runs over all nearest neighbor pairs
on the square lattice.
The doubled independent Ising model can be regarded as a decoupling point
$L=0$ of the Ashkin-Teller model.
Of course this is usually an unnecessary complication when one focuses on
the Ising model.
However, in the present problem it seems necessary to consider
the doubled Ising model as a special case of the Ashkin-Teller model,
This will become manifest in Sec.~\ref{subsec:spincorr},
where two-spin correlation functions are discussed.

The critical Ashkin-Teller model is identified with a $c=1$ conformal
field theory. Precisely speaking, it is not a simple free boson but
the $Z_2$ orbifold of the free boson.
(For a review, see Ref.~\cite{Ginsparg:LesHouches}.)
We take the normalization of the free boson field so that
the Lagrangian is given by
\begin{equation}
  {\cal L} = \frac{1}{2 \pi} (\partial_{\mu} \varphi)^2 .
\label{eq:bosonLag}
\end{equation}
The orbifold theory depends on a continuous parameter:
the compactification radius $r$ of the free boson
(i.e. we make the identification $\varphi \sim \varphi + 2 \pi r$).
The decoupling point (doubled Ising model), which
is relevant in our defect problem, corresponds to $r=1$.
The boundary condition (or the boundary state, see below) in the
present problem can be identified in two (presumably equivalent) ways:
in terms of Ashkin-Teller boundary states and in terms of the orbifold
free boson.
We first describe the former identification, which will be useful in
the calculation of the spin correlation function.
We will discuss the orbifold boundary state in the next section.

A boundary of the system is described by a boundary state,
if we exchange space and time so that the ``(imaginary) time''
is orthogonal to the boundary.
In particular, when the boundary is conformally invariant,
the boundary state $| B \rangle$ must satisfy the condition
\begin{equation}
\label{eq:Ishibashicond}
 (  L_n^{(P)} - \bar{L}_{-n}^{(P)} ) | B \rangle  =0,
\end{equation}
where $L^{(P)}$ and $\bar{L}^{(P)}$ are the bulk Virasoro
generators~\cite{Ishibashi:state,Cardy:fusion}.
A solution to this equation is given by the Ishibashi state
\begin{equation}
  | \Delta \rangle \rangle = \sum_{N} | \Delta, N \rangle \otimes | \overline{\Delta, N} \rangle
\end{equation}
where $| \Delta, N \rangle$ is a normalized $N$-th descendant of the primary
field of weight $\Delta$, and the summation is over all descendants.
In this paper, $ | \Delta \rangle \rangle$ denotes the Ishibashi state
constructed from the primary state with the weight $\Delta$.
A linear combination of Ishibashi states also satisfy conformal
invariance.
In general, the boundary state consists of several Ishibashi states.
The corresponding primary weights are included in
bulk operator contents.
While there are bulk operator with spin,
only the spinless primaries ($ \Delta = \bar{\Delta} $)
are relevant in the boundary state, as is seen from the structure
of the Ishibashi states.

The partition function and the operator content of the critical
Ashkin-Teller model is studied by Yang~\cite{Yang:ZofAT}.
We summarize the spinless primaries
at the decoupling point (doubled Ising model)
in Table~\ref{tab:ATops}.
Hereafter we will denote
the critical Ashkin-Teller model at the decoupling point
simply as ``Ashkin-Teller model''. 

\begin{table}[htbp]
  \begin{center}
    \leavevmode
\begin{tabular}{c|c|l|l}
$\Delta = \bar{\Delta} $  & Multiplicity & Degeneracy & Identification \\
\hline
\hline
$n^2$                 &  1 & [degenerate ] &
          $I_1 \cdot I_2$ ($n=0$); $\epsilon_1 \cdot \epsilon_2$ ($n=1$) \\
$\frac{(n+1)^2}{2}$   & 2 & [non-degenerate ]
           & $\epsilon_1 \cdot I_2, I_1 \cdot \epsilon_2 $ ($n=0$) \\
$\frac{(2n+1)^2}{8}$  & 1 & [non-degenerate ]  
           & $\sigma_1 \cdot \sigma_2$ ($n=0$) \\
$\frac{(2n+1)^2}{16}$ &  2 & [non-degenerate ] 
           & $\sigma_1 \cdot I_2 , I_1 \cdot \sigma_2$  ($n=0$) \\
\end{tabular}
\caption{The spinless primary operators in the critical Ashkin-Teller model
(at the decoupling point).
$n=0 , 1, 2 , \ldots$
and the ``degeneracy'' means the degeneracy in terms of
the $c=1$ Virasoro representation (not the multiplicity of the operators).
}
    \label{tab:ATops}
  \end{center}
\end{table}

A primary state $| I \rangle$ of the
Ising model satisfies the condition
\begin{equation}
  L_n^{\rm Ising} | I \rangle = 0 \;\;\;\; ( n > 0)
\end{equation}
where $L_n^{\rm Ising}$ is a Virasoro generator of the Ising model.
A primary state of the Ashkin-Teller model
is defined by the condition
\begin{equation}
  ( L_n^{\rm Ising 1} + L_n^{\rm Ising 2} ) | AT \rangle =0
\;\;\;\; (n >0)
\end{equation}
Thus a tensor product of two Ising primaries is a primary
(or a sum of primaries) of the Ashkin-Teller model, but
the reverse is not always true.
In fact, there is an infinite number of primaries in the Ashkin-Teller model,
while only three primaries are present in the Ising model:
the identity operator $I$, the energy density $\epsilon$ and the
spin operator $\sigma$. 
For example, the second set of the above spinless primaries
($\Delta = \bar{\Delta} = (n+1)^2/2$)
is identified as follows. There are two primaries for each dimension.
For $n$ even ($\Delta = \bar{\Delta} = (2m+1)^2/2$),
the two Ashkin-Teller primaries are $\epsilon_1 \cdot I_2$ and
$I_1 \cdot \epsilon_2$ ($n=0$) or corresponding
Ising descendants ($n \geq 1$).
We will distinguish them as $| \frac{(2m+1)^2}{2} , 1 \rangle \rangle$
and $| \frac{(2m+1)^2}{2} ,2 \rangle \rangle$.
For $n$ odd ($\Delta = \bar{\Delta} = 2m^2$),
the two primaries are Ising descendants of
$I_1 \cdot I_2$ and $\epsilon_1 \cdot \epsilon_2$.
We will distinguish corresponding Ishibashi states as
$| 2m^2 , II \rangle \rangle \equiv | 2m^2 , 1 \rangle \rangle$ and
$| 2m^2 , \epsilon \epsilon \rangle \rangle \equiv | 2m^2 , 2 \rangle \rangle$.

Let us write the tensor product of the Ising Ishibashi states
as Ashkin-Teller Ishibashi states.
For example, we consider the Ising Ishibashi state $| I \rangle \rangle$
of the identity operator $I$. (Note that it is not a physically allowed
boundary state of the Ising model without superposition
with other Ishibashi states~\cite{Cardy:fusion}.)
The partition function on a strip of width $1/2$ and length $\beta$
with two boundaries specified by the Ishibashi state $| I \rangle \rangle$
is given by the $c=1/2$ Virasoro character~\cite{Cardy:bceffect}
\begin{equation}
  \chi_{I} (\tilde{q})  =
  \frac{1}{2 \eta(\tilde{q})} [ \vartheta_3(\tilde{q}) + \vartheta_4(\tilde{q}) ] 
\label{eq:chIdIsing}
\end{equation}
where $\tilde{q}$ is defined as
\begin{equation}
\label{eq:deftildeq}
  \tilde{q} = e^{- 2 \pi / \beta} .
\end{equation}

Let us consider the Ashkin-Teller
boundary state defined by the tensor product of the Ising Ishibashi state
$| I_1 \rangle \rangle \otimes | I_2 \rangle \rangle$.
The partition function of the Ashkin-Teller model on the strip with
the corresponding boundary condition on both boundaries is given
by the square of eq.~(\ref{eq:chIdIsing}).
Regarding the Ashkin-Teller model as a $c=1$ conformal field theory,
it should be possible to express it in terms of
$c=1$ Virasoro characters.
The $c=1$ Virasoro characters~\cite{Kac:lec94} for the primary weight $h$
are given by
\begin{equation}
  \chi_h(q) = \frac{q^h}{\eta(q)},
\label{eq:c1character}
\end{equation}
if $h \neq n^2/4$.
When $ h = n^2/4$, the representation is degenerate and
the character $\chi_h$ is rather given by
\begin{equation}
  \chi_h(q) = \frac{q^{n^2/4} - q^{(n+2)^2/4}}{\eta(q)} .
\end{equation}
The partition function on the strip is expressed as
\begin{equation}
\chi_I(\tilde{q})^2 = \sum_{n=1}^{\infty} \chi_{2n^2}(\tilde{q}) +
                      \sum_{n=0}^{\infty} \chi_{(2n)^2} (\tilde{q}) .
\end{equation}
Thus the boundary state is identified in terms of Ashkin-Teller Ishibashi
states as
\begin{equation}
| I_1 \rangle \rangle \otimes | I_2 \rangle \rangle =
   \sum_{n=1}^{\infty} | 2 n^2, II \rangle \rangle +
   \sum_{n=1}^{\infty} | (2n)^2 \rangle \rangle .
\end{equation}
Similarly, we identify other tensor products as follows.
\begin{eqnarray}
  | \epsilon_1 \rangle \rangle \otimes | \epsilon_2 \rangle \rangle &=&
        \sum_{n=1}^{\infty} | 2n^2, \epsilon \epsilon \rangle \rangle +
        \sum_{n=0}^{\infty} | (2n+1)^2 \rangle \rangle 
\\
 | \sigma_1 \rangle \rangle \otimes | \sigma_2 \rangle \rangle &=&
        \sum_{n=0}^{\infty} | \frac{(2n+1)^2}{8} \rangle \rangle
\\
  | I_1 \rangle \rangle \otimes | \epsilon_2 \rangle \rangle &=&
        \sum_{n=0}^{\infty} | \frac{(2n+1)^2}{2} , 2 \rangle \rangle
\\
  | \epsilon_1 \rangle \rangle \otimes | I_2 \rangle \rangle &=&
        \sum_{n=0}^{\infty} | \frac{(2n+1)^2}{2} , 1 \rangle \rangle
\\
  | I_1 \rangle \rangle \otimes | \sigma_2 \rangle \rangle &=&
   \sum_{n \equiv 0,3 ({\rm mod}4)} | \frac{(2n+1)^2}{16} , 2 \rangle \rangle
\\
  | \sigma_1 \rangle \rangle \otimes | I_2 \rangle \rangle &=&
   \sum_{n \equiv 0,3 ({\rm mod}4)} | \frac{(2n+1)^2}{16} , 1 \rangle \rangle
\\
  | \epsilon_1 \rangle \rangle \otimes | \sigma_2 \rangle \rangle &=&
   \sum_{n \equiv 1,2 ({\rm mod} 4)} | \frac{(2n+1)^2}{16} , 2 \rangle \rangle
\\
  | \sigma_1 \rangle \rangle \otimes | \epsilon_2 \rangle \rangle &=&
   \sum_{n \equiv 1,2 ({\rm mod} 4)} | \frac{(2n+1)^2}{16} , 1 \rangle \rangle
\end{eqnarray}

The boundary states of the Ising model
correspond to the free, up-spin and down-spin
boundary conditions are given by~\cite{Cardy:fusion}
\begin{eqnarray}
  | f \rangle &=& | I \rangle \rangle - | \epsilon \rangle \rangle
\\
  | \uparrow \rangle &=&
         \frac{1}{\sqrt{2}}(| I \rangle \rangle + | \epsilon \rangle \rangle )
        +  \frac{1}{2^{1/4}} | \sigma \rangle \rangle .
\\
  | \downarrow \rangle &=&
         \frac{1}{\sqrt{2}}(| I \rangle \rangle + | \epsilon \rangle \rangle )
        -  \frac{1}{2^{1/4}} | \sigma \rangle \rangle .
\end{eqnarray}
The $b=0$ defect in our model is described by the tensor product
of the free boundary state of the Ising model as
\begin{eqnarray}
  | ff \rangle &=& \left( | I_1 \rangle \rangle - | \epsilon_1 \rangle \rangle
                   \right)
          \otimes \left( | I_2 \rangle \rangle - | \epsilon_2 \rangle \rangle
                  \right)
\nonumber \\
&=&
\sum_{j=1}^{2} \sum_{n=1}^{\infty}
   \cos{\frac{\pi n}{2}} | \frac{n^2}{8},j \rangle \rangle
+ \sum_{n=0}^{\infty} | n^2 \rangle \rangle.
\label{eq:ff}
\end{eqnarray}
Similarly, the fixed spin boundary states of the Ashkin-Teller model
are given by
\begin{eqnarray}
  | \uparrow \uparrow \rangle &=& 
    | \uparrow_1 \rangle \otimes | \uparrow_2 \rangle 
\nonumber \\
&=&
\frac{1}{2} \sum_{j=1}^{2} \sum_{n=1}^{\infty}
             | \frac{n^2}{2},j \rangle \rangle
    + \frac{1}{2} \sum_{n=0}^{\infty} | n^2 \rangle \rangle
\nonumber \\
&&
+ \frac{1}{\sqrt{2}} \sum_{n=0}^{\infty} | \frac{(2n+1)^2}{8} \rangle \rangle
+ \frac{1}{2^{3/4}} \sum_{j=1}^{2} \sum_{n=0}^{\infty} 
        | \frac{(2n+1)^2}{16}, j \rangle \rangle
\label{eq:upup}
\\
| \downarrow \downarrow \rangle &=& 
\frac{1}{2} \sum_{j=1}^{2} \sum_{n=1}^{\infty}
       | \frac{n^2}{2},j \rangle \rangle
    + \frac{1}{2} \sum_{n=0}^{\infty} | n^2 \rangle \rangle
         \nonumber \\
&&
+ \frac{1}{\sqrt{2}} \sum_{n=0}^{\infty} | \frac{(2n+1)^2}{8} \rangle \rangle
- \frac{1}{2^{3/4}} \sum_{j=1}^{2} \sum_{n=0}^{\infty} 
        | \frac{(2n+1)^2}{16}, j \rangle \rangle
\\
| \uparrow \downarrow \rangle &=& 
\frac{1}{2} \sum_{j=1}^{2} \sum_{n=1}^{\infty}
                 | \frac{n^2}{2},j \rangle \rangle
    + \frac{1}{2} \sum_{n=0}^{\infty} | n^2 \rangle \rangle
    - \frac{1}{\sqrt{2}}  \sum_{n=0}^{\infty} 
        | \frac{(2n+1)^2}{8} \rangle \rangle
\nonumber \\
&&
- \frac{1}{2^{3/4}} \sum_{j=1}^{2} (-1)^j \sum_{n=0}^{\infty} 
        | \frac{(2n+1)^2}{16}, j \rangle \rangle
\\
| \downarrow \uparrow \rangle &=& 
\frac{1}{2} \sum_{j=1}^{2} \sum_{n=1}^{\infty}
                 | \frac{n^2}{2},j \rangle \rangle
    + \frac{1}{2} \sum_{n=0}^{\infty} | n^2 \rangle \rangle
    - \frac{1}{\sqrt{2}}  \sum_{n=0}^{\infty} 
        | \frac{(2n+1)^2}{8} \rangle \rangle
\nonumber \\
&&
+ \frac{1}{2^{3/4}} \sum_{j=1}^{2} (-1)^j \sum_{n=0}^{\infty} 
        | \frac{(2n+1)^2}{16}, j \rangle \rangle
\label{eq:downup}
\end{eqnarray}
Furthermore, there are tensor products of free and fixed boundary
states:
\begin{eqnarray}
  | \uparrow f \rangle &=&
    \frac{1}{2} \sum_{n=1}^{\infty} | \frac{n^2}{2} , A \rangle \rangle
+ \frac{1}{2} (-1)^n | n^2 \rangle \rangle
\nonumber \\
&&+ \frac{1}{2^{1/4}} \sum_{n=0}^{\infty} 
 (-1)^{\frac{n(n+1)}{2}}   | \frac{(2n+1)^2}{16} , 1 \rangle \rangle
\\
 | \downarrow f \rangle &=& 
    \frac{1}{2} \sum_{n=1}^{\infty} | \frac{n^2}{2} , A \rangle \rangle
+ \frac{1}{2} (-1)^n | n^2 \rangle \rangle
\nonumber \\
&& - \frac{1}{2^{1/4}} \sum_{n=0}^{\infty} 
 (-1)^{\frac{n(n+1)}{2}}   | \frac{(2n+1)^2}{16} , 1 \rangle \rangle
\\
| f \uparrow \rangle &=&
    \frac{1}{2} \sum_{n=1}^{\infty} (-1)^n | \frac{n^2}{2} , A \rangle \rangle
+ \frac{1}{2} (-1)^n | n^2 \rangle \rangle
\nonumber \\
&&+ \frac{1}{2^{1/4}} \sum_{n=0}^{\infty} 
 (-1)^{\frac{n(n+1)}{2}}   | \frac{(2n+1)^2}{16} , 2 \rangle \rangle
\\
| f \downarrow \rangle &=&
    \frac{1}{2} \sum_{n=1}^{\infty} (-1)^n | \frac{n^2}{2} , A \rangle \rangle
+ \frac{1}{2} (-1)^n | n^2 \rangle \rangle
\nonumber \\
&& - \frac{1}{2^{1/4}} \sum_{n=0}^{\infty} 
 (-1)^{\frac{n(n+1)}{2}}   | \frac{(2n+1)^2}{16} , 2 \rangle \rangle
\end{eqnarray}
where
\begin{equation} 
| \frac{n^2}{2}, A \rangle  \equiv
\frac{1}{\sqrt{2}} 
\left[
 | \frac{n^2}{2}, 1 \rangle - | \frac{n^2}{2},2 \rangle 
\right] .
\label{eq:ATBSA}
\end{equation}
It is natural to think that the combination
\begin{equation}
  | \uparrow \uparrow \rangle + | \downarrow \downarrow \rangle =
    \sum_{j=1}^2 \sum_{n=1}^{\infty} | \frac{n^2}{2} , j \rangle \rangle
  + \sqrt{2} \sum_{n=0}^{\infty} | \frac{(2n+1)^2}{8} \rangle \rangle
  + \sum_{n=0}^{\infty} | n^2 \rangle \rangle
\label{eq:upupdndn}
\end{equation}
corresponds to the infinitely ferromagnetic defect $b \rightarrow \infty$
together with the anisotropic limit $ K_2 \rightarrow 0$,
namely $\alpha = - \frac{\pi }{4}$.
(The overall normalization is determined from the partition function.
See below.)
Similarly,
\begin{equation}
  | \uparrow \downarrow \rangle + | \downarrow \uparrow \rangle =
    \sum_{j=1}^2 \sum_{n=1}^{\infty} | \frac{n^2}{2} , j \rangle \rangle
  - \sqrt{2} \sum_{n=0}^{\infty}
          | \frac{(2n+1)^2}{8} \rangle \rangle
  + \sum_{n=0}^{\infty} | n^2 \rangle \rangle
\label{eq:updndnup}
\end{equation}
corresponds to the infinitely antiferromagnetic defect $b \rightarrow - \infty$
with $K_2 \rightarrow 0$.

Thus we have obtained the boundary state for three special types
of defect line.
Now we make the following
assumption about the boundary state for general strength
of the defect, generalizing~(\ref{eq:ff}),(\ref{eq:upupdndn}) and
(\ref{eq:updndnup}).

\begin{quotation}
\subsection*{Assumption}
{\em
The boundary state $| B \rangle$ is given by
\begin{equation}
  | B (\varphi_0) \rangle = \sum_{n=0}^{\infty} | n^2 \rangle \rangle
        + \sqrt{2} \sum_{n=1}^{\infty} \cos{( n \varphi_0)}
                       | \frac{n^2}{8} , S \rangle \rangle .
\label{eq:bstateAT}
\end{equation}
Here $\varphi_0$ is given by
\begin{equation}
  \varphi_0 = \frac{\pi}{4} + \frac{\alpha}{2},
\label{eq:phi0def}
\end{equation}
using $\alpha$ defined in~(\ref{eq:alphadef}). 
$| n^2/8 ,S \rangle \rangle$ is defined as
\begin{equation}
  | \frac{n^2}{8} , S \rangle \rangle =
   \left\{ \begin{array}{ll}
      2^{-1/2}
      \left[ | (2k)^2/8 ,1 \rangle \rangle + 
          | (2k)^2/8,2 \rangle \rangle \right]
   & ( n = 2k ) 
   \\
   | (2k+1)^2/8 \rangle \rangle & ( n = 2k+1 )  
   \end{array} \right.    
\label{eq:bstateSym}
\end{equation}
}
\end{quotation}

We note that the boundary state contains only the combination
$|n^2/2,S \rangle \rangle$ of Ishibashi states, which is symmetric
with respect to the interchange of the two Ising layers.
This is presumably a consequence of the reflection symmetry
about the defect line in the unfolded picture.
For the boundary corresponding to the defect line, we take
$\alpha = \alpha(b)$. For the other boundary, we take $\varphi_0 = \pi/4$
because there is no defect ($b=1$).
We summarize the correspondence between the defect strength
and the parameter $\varphi_0$ in Table~\ref{tab:phi0}.

\begin{table}[htbp]
  \begin{center}
    \leavevmode
    \begin{tabular}{c|c|l|l}
      $\varphi_0$ & $b$ & Condition & Description\\
\hline
      $0$         & $\infty$ &  $K_2 \rightarrow 0$
                        & Infinitely ferromagnetic limit\\
      $\pi/4$     & $1$      & (none) & No defect / periodic \\
      $\pi/2$     & $0$ & (none) & Free boundary \\
      $3 \pi/4$  & $-1$ & (none) & Antiferromagnetic defect / antiperiodic \\
      $\pi$    & $- \infty$ & $K_2 \rightarrow 0$
                  & Infinitely antiferromagnetic limit
    \end{tabular}
    \caption{The value of $\varphi_0$ for several strengths of the defect}
    \label{tab:phi0}
  \end{center}
\end{table}
We can calculate the partition function on the strip from the
boundary state.
Using the transfer matrix $H^{(P)}$ in the
direction orthogonal to the boundaries,
\begin{eqnarray}
Z &=&
  \langle B(\frac{\pi}{4}) | e^{- H^{(P)}/(2 \beta)} | B(\varphi_0) \rangle
\nonumber \\
&=& 
\sum_{n=0}^{\infty} \chi_{n^2}(\tilde{q})
+
2 \sum_{n=1}^{\infty} \cos{(n \varphi_0 )} \cos{(n \pi/4 )}
            \tilde{\chi}_{n^2/8}(\tilde{q})
\nonumber \\
&=&
Z(\varphi_0-\frac{\pi}{4}) + Z(\varphi_0 + \frac{\pi}{4})
\end{eqnarray}
where
\begin{eqnarray}
  Z(\xi) &=& \frac{1}{2 \eta(\tilde{q})}
               \vartheta_3(e^{i \xi}, \tilde{q}^{1/4})
\nonumber \\
&=& \frac{1}{\eta(q)} e^{-  \beta \xi^2/\pi}
               \vartheta_3(e^{-4 \xi \beta},q^4)
\nonumber \\
&=& 
\frac{1}{2 \eta(q)} e^{-  \beta \xi^2/\pi}
       [ \vartheta_3(e^{-2 \xi \beta},q) + \vartheta_4(e^{-2 \xi \beta},q)]
\label{eq:Zofphi}
\end{eqnarray}
This actually agrees with the
partition function of the model obtained in~(\ref{eq:totalZ}).

We note that, while Henkel {\it et al.}~\cite{Henkel87,Henkel89}
identified the partition function with a
$c=1$ Virasoro algebra empirically, 
it is a natural consequence from our ``folding'' approach.
They further studied the system with many parallel
defect lines~\cite{Henkel89}.
For example, for the system with two defects,
they found that the spectum can be described in terms of
a Virasoro algebra if the location of the defect lines
is commensurate, namely the distance between two defects
is an integral multiple of $1/n$ of the system size, where
$n$ is an integer.
The central charge in this case is given by $c=n$ when $n$ is odd
and $c=n/2$ if $n$ is even.
This can be also naturally understood in terms of multiple folding,
i.e. we fold the system many times so that we have $2c$ layers of the
Ising model, and defect lines are placed only at the boundaries.
In principle the boundary conformal field theory
for the corresponding central charge
could be useful for analysis of the commensurate multiple defects.
Such an analysis is however beyond the scope of the present
paper.

\section{$Z_2$ orbifold of free boson and the boundary state}
\label{sec:orbstate}

We turn to the second identification of the boundary state
in terms of the free boson.
This is useful in calculation of some correlation functions,
as we will see in the Section~\ref{sec:Dcorr}.
Moreover, it gives a more transparent understanding of boundary states
and also enables us to find a new universality class of defect lines.

\subsection{Boundary states of the free boson}

In order to make our paper self-contained, here we give a brief review on the
boundary states of the free boson.
We consider the free boson (before the orbifolding) defined
by the Lagrangian density
\begin{equation}
  {\cal L} = \frac{1}{2 \pi} ( \partial_{\mu} \varphi )^2
\end{equation}
where $\varphi = \varphi(\sigma, t)$ and $0 \leq \sigma \leq \beta$.
$\varphi$ is compactified with the compactification radius $r$,
namely $\varphi \sim \varphi + 2 \pi r$.
From the Lagrangian, we can derive the equation of motion
\begin{equation}
\label{eq:bosoneqm}
  \partial^2 \varphi =0
\end{equation}
and the canonical commutation relation
\begin{equation}
\label{eq:bosoncomm}
  [ \varphi(\sigma, t) , \Pi(\sigma' , t) ] = i \delta ( \sigma - \sigma' )
\end{equation}
where $\Pi = \dot{\varphi} / \pi$.
We impose the
periodic boundary condition in ``space'' (parallel to the boundary
direction) $\varphi(\sigma,t) \sim \varphi(\sigma + \beta, t)$.

We can determine the mode expansion of the boson field from
eqs.~(\ref{eq:bosoneqm}) and (\ref{eq:bosoncomm}) as follows:
\begin{eqnarray}
\lefteqn{  \varphi(\sigma,t) =} \nonumber \\
&& \hat{x} + \frac{2 \pi}{\beta} r w \sigma + \frac{\pi}{\beta} \hat{p} t
   \nonumber \\
&& + \frac{1}{2}  \sum_{n = 1}^{\infty} [
   \frac{a_n}{\sqrt{n}} e^{ - in( \sigma + t) \frac{2 \pi}{\beta}}
+ \frac{a^{\dagger}_n}{\sqrt{n}} e^{ in( \sigma + t) \frac{2 \pi}{\beta}}
           ]
\nonumber \\
&& + \frac{1}{2}  \sum_{n = 1}^{\infty} [
   \frac{\tilde{a}_n}{\sqrt{n}} e^{ in( \sigma - t) \frac{2 \pi}{\beta}}
+ \frac{\tilde{a}^{\dagger}_n}{\sqrt{n}}
e^{ - in( \sigma - t) \frac{2 \pi}{\beta}} ]
\end{eqnarray}
where $w$ is an integer (winding number) allowed by the angular nature
of the boson field.
The operators satisfies the commutation relations
\begin{eqnarray}
 {[} \hat{x} , \hat{p} ] & = & i  \\
 {[} a_n , {a}^{\dagger}_{m}] =
 {[} \tilde{a}_n , \tilde{a}^{\dagger}_m ]
 &=& \delta_{nm}
\label{eq:osccomm}
\end{eqnarray}
and the other commutators vanish.
Since the constant mode $\hat{x}$ is also compactified as
$\hat{x} \sim \hat{x} + 2 \pi r$, the conjugate momentum $\hat{p}$
is quantized to an integer multiple of $1/r$.

The boson field $\varphi$ can be decomposed into chiral components
$\varphi_L$ and $\varphi_R$ as
\begin{eqnarray}
  \varphi_L ( x^+) &=&
  \frac{\hat{x}}{2} + \frac{\pi}{\beta}(rw + \frac{\hat{p}}{2}) x^+
\nonumber \\
&& + \frac{1}{2}  \sum_{n = 1}^{\infty} [
   \frac{a_n}{\sqrt{n}} e^{ - in x^+ \frac{2 \pi}{\beta}}
+ \frac{a^{\dagger}_n}{\sqrt{n}} e^{ in x^+ \frac{2 \pi}{\beta}}
           ] 
\\
  {\varphi}_{R} ( x^- ) &=&
  \frac{\hat{x}}{2} + \frac{\pi}{\beta}(- rw + \frac{\hat{p}}{2}) x^-
\nonumber \\
&& + \frac{1}{2}  \sum_{n = 1}^{\infty} [
   \frac{\tilde{a}_n}{\sqrt{n}} e^{ - in x^- \frac{2 \pi}{\beta}}
+ \frac{\tilde{a}^{\dagger}_n}{\sqrt{n}} e^{ in x^- \frac{2 \pi}{\beta}}
           ]
\end{eqnarray}
where $x^{\pm} = t \pm \sigma$.

We make a mode expansion of the energy-momentum tensor
$T(x^+) = \frac{1}{\pi} (\partial_+ \varphi_L )^2$ and
$\bar{T}(x^-) = \frac{1}{\pi} (\partial_- \varphi_R)^2$
as
\begin{eqnarray}
  T(x^+) &=& \frac{2 \pi}{\beta^2} \sum_{m = -\infty}^{\infty}
           L_m e^{- i m x^+ \frac{2 \pi}{\beta}}
\\
  T(x^-) &=& \frac{2 \pi}{\beta^2} \sum_{m = - \infty}^{\infty}
         \bar{L}_m e^{- i m x^- \frac{2 \pi}{\beta}} .
\end{eqnarray}
$L_m$ and $\bar{L}_m$, which are (bulk) Virasoro generator,
are given by
\begin{eqnarray}
  L_m = \frac{1}{2} \sum_{n = - \infty}^{\infty}
           : \alpha_{m-n} \alpha_n : - \frac{1}{12} \delta_{m0}
\\
  \bar{L}_m = \frac{1}{2} \sum_{n = - \infty}^{\infty}
      : \tilde{\alpha}_{m-n} \tilde{\alpha}_n : - \frac{1}{12} \delta_{m0},
\end{eqnarray}
where $\alpha_m$ and $\tilde{\alpha}_m$ are defined as
\begin{eqnarray}
  \alpha_n &=&
    \left\{  \begin{array}{lr}
           - i \sqrt{n} a_n & ( n > 0)  \\
           ( rw + \frac{\hat{p}}{2} ) & ( n=0 ) \\
            i \sqrt{n} a^{\dagger}_{-n} & ( n < 0 ) 
           \end{array} \right.
\\
  \tilde{\alpha}_n &=&
    \left\{  \begin{array}{lr}
           - i \sqrt{n} \tilde{a}_n & ( n > 0)  \\
           ( - rw + \frac{\hat{p}}{2} ) & ( n=0 ) \\
            i \sqrt{n} \tilde{a}^{\dagger}_{-n} & ( n < 0 ) 
           \end{array} \right.
\end{eqnarray}
and $::$ denotes the normal ordering
\begin{equation}
  : \alpha_n \alpha_m : = \left\{ \begin{array}{lr}
                 \alpha_n \alpha_m & ( n \leq m ) \\
                 \alpha_m \alpha_n & ( n > m )
          \end{array} \right.
.
\end{equation}
The Hamiltonian for the free boson with the periodic boundary
condition in the spatial direction of length $\beta$ is given by
\begin{equation}
 H^{(P)}_{\beta} = \frac{2 \pi}{\beta} \left[
               (rw)^2 + (\frac{\hat{p}}{2})^2 
              + \sum_n n a^{\dagger}_n a_n
              + \sum_n n \tilde{a}^{\dagger}_n \tilde{a}_n
              - \frac{1}{12} \right] .
\end{equation}
A conformally invariant boundary state must satisfy the Ishibashi
condition~(\ref{eq:Ishibashicond}).
In the present case, this condition is satisfied, if the boundary
state $| X \rangle$ satisfies 
\begin{equation}
  ( \alpha_m \pm \tilde{\alpha}_{-m} ) | X \rangle = 0
\end{equation}
for any integer $m$. (The sign $\pm$ should be common for all $m$.)
Let us first consider
\begin{equation}
\label{eq:IshibashiDir}
 ( \alpha_m + \tilde{\alpha}_{-m} ) | X \rangle = 0 .
\end{equation}
The conditions for $m \neq 0$ are satisfied by
\begin{equation}
  | X \rangle = 
       \exp{[ - \sum_{n=1}^{\infty} a^{\dagger}_n \tilde{a}^{\dagger}_n ] }
       | \mbox{vac} \rangle,
\end{equation}
where $| \mbox{vac} \rangle$ is an oscillator vacuum
which satisfies
$ a_n | \mbox{vac} \rangle = \tilde{a}_n | \mbox{vac} \rangle = 0$.
An oscillator vacuum is characterized by two integers: $w$ and
$k = r \hat{p}$. (Recall that $\hat{p}$ is quantized due to
the compactification.)
Hereafter $| ( w, k) \rangle$ denotes the oscillator vacuum with
zero-mode parameters $w$ and $k$.
The remaining condition for $m=0$ requires $w=0$.
Thus a boundary state which satisfies eq.~(\ref{eq:IshibashiDir}) is given by
\begin{equation}
  \sum_k C_k 
   \exp{[ - \sum_{n=1}^{\infty} a^{\dagger}_n \tilde{a}^{\dagger}_n ] }
   | ( 0, k ) \rangle , 
\end{equation}
where $C_k$ are constants.
Besides the Ishibashi condition, there is a consistency
condition found by Cardy~\cite{Cardy:fusion}.
We can calculate the partition function on a strip with
two boundary conditions.
If we exchange the space and time and regard
the direction parallel to the boundary as time coordinate,
the partition function should be expressed as a sum of Virasoro
characters with integer coefficients. Namely, the partition function
with the boundary conditions $A$ and $B$ on the two sides
should take the form
\begin{equation}
  Z = \sum_{h} n^h_{AB} \chi_h(q) 
\label{eq:consistency}
\end{equation}
where $n^h_{AB}$ are non-negative integers and $\chi_h(q)$ is
the Virasoro character for the primary weight $h$.
If the two boundary conditions are the same,
the partition function shows the boundary operator content for
the boundary.
In particular,
there should be exactly one dimension-zero character corresponding
to the identity operator: $n^0_{AA} = 1$.
These requirement comes from the radial quantization on a half-plane
and the conformal mapping to the strip.

A solution to the consistency condition is given by
\begin{equation}
  C_k = \frac{1}{\sqrt{2 r}} e^{- i k \varphi_0 /r},
\end{equation}
where $\varphi_0$ is a constant.
It gives a one-parameter family of boundary states
\begin{equation}
\label{eq:DirBS}
  | D ( \varphi_0) \rangle =
  \frac{1}{\sqrt{2 r}} 
   \sum_{k= - \infty}^{\infty} e^{- i k \varphi_0 /r}
   \exp{[ - \sum_{n=1}^{\infty} a^{\dagger}_n \tilde{a}^{\dagger}_n ] }
   | ( 0, k ) \rangle , 
\end{equation}
Below we show that this state satisfies the consistency condition.
Let us assume the width of the strip is $1/2$ (and the length is $\beta$).
The partition function of the strip for the above boundary condition
with parameters $\varphi_0$ and $\varphi_0'$ at two boundaries is
\begin{eqnarray}
   Z_r(\Delta \varphi_0) & \equiv &   
  \langle D ( \varphi_0 ) | e^{-H^{(P)}_{\beta}/2} | D( \varphi_0') \rangle 
\\ \nonumber
&=& 
\frac{1}{2 r} \frac{1}{\eta ( \tilde{q} )}
             \vartheta_3 ( e^{i \Delta \varphi_0 /r} , \tilde{q}^{1/4 r^2}),
\label{eq:Zrboson}
\end{eqnarray}
where $\tilde{q}$ is defined in eq.~(\ref{eq:deftildeq}) and
$\Delta \varphi_0 = \varphi_0 - \varphi_0'$.
Here $1 / \eta (\tilde{q}) $ is given by the summation over the
oscillator states and $\vartheta_3$ part comes from the summation over
zero-mode quantum number $k$.
We express the above partition function as a function of $q$
defined in eq.~(\ref{eq:qandw}). This is achieved by
a modular transformation of $\eta$ and $\vartheta$ function.
The result is
\begin{equation}
  Z_r( \Delta \varphi_0 )  = \frac{1}{\eta( q )} 
           q^{-(\Delta \varphi_0 / \pi)^2} 
           \vartheta_3 ( e^{-4 r \Delta \varphi_0 \beta} , q^{4 r^2}) ,
\end{equation}
This actually is a sum of $c=1$ Virasoro characters with 
non-negative integer coefficients.
In particular, it gives the boundary operator content when
the two boundary states are the same ($\Delta \varphi_0 = 0$).
In this case, the partition function contains one dimension-zero
character.
Thus Cardy's consistency conditions are satisfied.
The boundary state~(\ref{eq:DirBS}) has a simple physical meaning:
we see that
\begin{equation}
  \varphi(\sigma, t=0) | D ( \varphi_0 ) \rangle
             = \varphi_0 | D (\varphi_0 ) \rangle ,
\end{equation} 
namely the boson field takes a fixed value $\varphi_0$ at the boundary.
Thus it corresponds to the Dirichlet boundary condition
$\varphi = \varphi_0$. 

We can discuss the other possibility
\begin{equation}
 ( \alpha_m - \tilde{\alpha}_{-m} ) | X \rangle = 0
\end{equation}
in a similar manner.
In this case, conditions for $m \neq 0$ implies
\begin{equation}
  | X \rangle = 
       \exp{[ + \sum_{n=1}^{\infty} a^{\dagger}_n \tilde{a}^{\dagger}_n ] }
       | \mbox{vac} \rangle,
\end{equation}
and that for $m=0$ requires $\hat{p} =0$.
A solution to Cardy's consistency condition is given by
\begin{equation}
\label{eq:NeuBS}
  | N ( \tilde{\varphi}_0) \rangle =
  \sqrt{r}
   \sum_{w= - \infty}^{\infty} e^{- 2 i r w \tilde{\varphi}_0}
   \exp{[ + \sum_{n=1}^{\infty} a^{\dagger}_n \tilde{a}^{\dagger}_n ] }
   | ( w, 0 ) \rangle .
\end{equation}
Again it has a simple physical interpretation: von Neumann boundary
condition ($\dot{\varphi} = \mbox{const.}$), or equivalently,
Dirichlet boundary condition on the dual field
$\tilde{\varphi} \equiv \varphi_L - \varphi_R = \tilde{\varphi}_0$.
Since the von Neumann boundary condition is the Dirichlet boundary
condition on the dual field, 
the amplitude between two von Neumann boundary states are given by
replacing the compactification radius by its dual $1/(2r)$,
and the parameters $\varphi_0$ and $\varphi_0'$ by 
$\tilde{\varphi}_0$ and $\tilde{\varphi}_0'$
in eq.~(\ref{eq:Zrboson}).
Thus the von Neumann boundary state is mutually consistent with
a von Neumann boundary state with any value of the parameter.
We note that, although the Dirichlet and Neumann boundary states contain
a continuous parameter, the operator content of the theory does not
depend on the parameter, as is seen from the amplitude with the
same boundary state.

The amplitude between the Dirichlet and the von Neumann boundary
states is given as
\begin{eqnarray}
  \langle D ( \varphi_0 ) | e^{-H^{(P)}_{\beta}/2}
              | N( \tilde{\varphi}_0) \rangle 
&=&
 \frac{1}{\sqrt{2}} \tilde{q}^{-1/24} \prod_{n=1}^{\infty} 
         \frac{1}{1+\tilde{q}^n}
\nonumber \\
&=& \frac{1}{\sqrt{2} \eta (\tilde{q})}{\vartheta_4(\tilde{q}^2)}
\nonumber \\
&=& \frac{1}{2 \eta (q)}{\vartheta_2(q^{1/2})} .
\end{eqnarray}
This is again a sum of $c=1$ Virasoro characters with non-negative
integer coefficients; they are mutually compatible.
We note that this Dirichlet-von Neumann amplitude depends
neither on $\varphi_0$ nor $\tilde{\varphi}_0$, because
only the zero winding number sector built on $|(0,0) \rangle$
contributes to the amplitude.

\subsection{Boundary states of the orbifold}
\label{sec:bstateorb}

Now we discuss the boundary states of the $Z_2$ orbifold,
which has a direct relevance to our problem.
Related discussion in string theory can be found in
Refs.~\cite{HarveyMinahan,IshibashiOnogi,PradisiSagnotti}.
The $Z_2$ orbifold of the free boson is constructed
from the free boson, identifying $\varphi \sim - \varphi$.
Thus a boundary state must be invariant under the transformation
\begin{equation}
  G: \varphi \rightarrow - \varphi .
\label{eq:invertphi}
\end{equation}
The simplest way to satisfy this requirement is to symmetrize
the free boson boundary state.
Such a boundary state constructed from the Dirichlet boundary
state of the free boson is
\begin{equation}
  | D_O (\varphi_0) \rangle = \frac{1}{\sqrt{2}}
       [ | D(\varphi_0) \rangle + | D( - \varphi_0) \rangle ] ,
\label{eq:bstateboson}
\end{equation}
where $ 0 < \varphi_0 < \pi r$.
The overall factor is determined from the consistency condition.
The partition function of the orbifold theory
for the Dirichlet condition
with parameters $\varphi_0$ and $\varphi_0'$ at the two boundaries
is given by
\begin{equation}
\langle D_O(\varphi_0) | e^{- H^{(P)}_{\beta}/2}
         | D_O (\varphi_0') \rangle
= Z_r(\varphi_0 - \varphi_0' ) + Z_r (\varphi_0 + \varphi_0' ) ,
\label{eq:Zphiphi}
\end{equation}
where $Z_r$ is defined in~(\ref{eq:Zrboson}). 
Thus the orbifold Dirichlet boundary state~(\ref{eq:bstateboson})
is mutually consistent boundary state
for any $0 < \varphi_0 < \pi r$.
The operator content for the boundary
state~(\ref{eq:bstateboson}) is obtained from the above
partition function with setting $\varphi_0' = \varphi_0$.
We note that, it contains $Z_r(2 \varphi_0)$ which depends
on the continuous parameter $\varphi_0$.
Thus, the boundary operator content of the orbiford theory
depends on the boundary value paramter $\varphi_0$.
On the other hand, the boundary operator content for
the Dirichlet boundary state~(\ref{eq:DirBS})
of the (unorbifolded) free boson does not depend on the
boundary value $\varphi_0$.
This is a consequence of the $U(1)$ symmetry
$\varphi \rightarrow \varphi + \mbox{const.}$ of the free boson.
The orbifolding breaks this symmetry and thus the spectrum of the
boundary operators can depend on $\varphi_0$.

Similarly, we can consider the orbifold version of the von Neumann
boundary state:
\begin{equation}
| N_O ( \tilde{\varphi}_0 \rangle \equiv
\frac{1}{\sqrt{2}}
\left[ | N(\tilde{\varphi}_0 \rangle + | N( - \tilde{\varphi}_0) \rangle 
\right]  ,
\label{eq:orbNeuBS}
\end{equation}
where $| N (\tilde{\varphi}_0) \rangle $ is given in~(\ref{eq:NeuBS})
and $ 0 < \tilde{\varphi}_0 < \pi / (2r)$.
It can be also shown that it is mutually consistent boundary
state, and also consistent with the orbifold Dirichlet
boundary state~(\ref{eq:bstateboson}).
From the analysis of the partition function,
the operator content of the state~(\ref{eq:orbNeuBS}) also
depends on $\tilde{\varphi}_0$.
 
The endpoints $\varphi_0 = 0, \pi r$ and $\tilde{\varphi}_0 = 0, \pi/(2r)$
are the fixed point of the transformation~(\ref{eq:invertphi}).
At those endpoints, the orbifold boundary state~(\ref{eq:bstateboson})
or~(\ref{eq:orbNeuBS}) contains two dimension-zero boundary operators
and therefore does not satisfy Cardy's consisntency condition.
On the other hand, another kind of Dirichlet/von Neumann boundary
state is possible at these endpoints.
 
Namely, due to the orbifold condition $\varphi \sim - \varphi$,
the antiperiodic boundary condition in terms of free boson
\begin{equation}
\varphi(\sigma + \beta , t) \sim - \varphi(\sigma , t)
\label{eq:bosonapbc}
\end{equation}
is allowed under the ``periodic'' boundary condition of the orbifold.
Let us call the subspace with the above boundary condition
the twisted sector.
The mode expansion of the free boson with the antiperiodic boundary
condition is given by
\begin{eqnarray}
\lefteqn{  \varphi(\sigma,t) =} \nonumber \\
&& \hat{x}
+ \frac{1}{2}  \sum_{n = 1}^{\infty} \left[
   \frac{b_n}{\sqrt{n - \frac{1}{2}}}
      e^{ - i (n - \frac{1}{2}) ( \sigma + t) \frac{2 \pi}{\beta}}
+ \frac{b^{\dagger}_n}{\sqrt{n - \frac{1}{2}}}
  e^{ i(n - \frac{1}{2})( \sigma + t) \frac{2 \pi}{\beta}}
           \right]
\nonumber \\
&& + \frac{1}{2}  \sum_{n = 1}^{\infty} \left[
   \frac{\tilde{b}_n}{\sqrt{n - \frac{1}{2}}}
   e^{ i(n - \frac{1}{2}) ( \sigma - t) \frac{2 \pi}{\beta}}
 + \frac{\tilde{b}^{\dagger}_n}{\sqrt{n - \frac{1}{2}}}
 e^{ - i(n - \frac{1}{2})( \sigma - t) \frac{2 \pi}{\beta}} 
  \right] ,
\end{eqnarray}
where $b_n$,$b^{\dagger}_n$,$\tilde{b}_n$ and $\tilde{b}^{\dagger}_n$
are the boson creation/annihilation operators obeying the commutation
relations similar to~(\ref{eq:osccomm}).
The Hamiltonian for the free boson with the antiperiodic boundary
condition in the spatial direction of length $\beta$ is given by
\begin{equation}
 H^{(A)}_{\beta} = \frac{2 \pi}{\beta} \left[
              \sum_n (n - \frac{1}{2}) b^{\dagger}_n b_n
              + \sum_n (n - \frac{1}{2}) \tilde{b}^{\dagger}_n \tilde{b}_n
              + \frac{1}{48} \right] .
\end{equation}

The essential point in the twisted sector is that
$\hat{x}$ can only take the fixed point values
$0$ or $\pi r$ due to the the antiperiodic
boundary condition~(\ref{eq:bosonapbc}).
Thus there are two independent oscillator vacua $|0 \rangle_T$
and $|\pi r \rangle_T$ corresponding to $\hat{x} = 0 , \pi r$,
in the twisted sector.
From them we can construct the Dirichlet boundary state
in the twisted sector.
Those twisted Dirichlet boundary states are given by
\begin{equation}
| D( \varphi_0 )_T \rangle =
  e^{ - \sum_{n=1}^{\infty} b^{\dagger}_n \tilde{b}^{\dagger}_n }
  | \varphi_0 \rangle_T  ,
\end{equation}
where $\varphi_0 = 0$ or $\pi r$ {\em only}.

The von Neumann boundary states in the twisted sector are less clear.
Since $\varphi$ is not fixed in the von Neumann boundary condition,
we assume that they are constructed on the oscillator vacua
$ | 0 \rangle_T \pm | \pi r \rangle_T$.
We will later confirm that this ansatz actually leads to mutually consistent
boundary states.
Then the twisted von Neumann boundary states are given by
\begin{eqnarray}
| N(0)_T \rangle &=&
  e^{ + \sum_{n=1}^{\infty} b^{\dagger}_n \tilde{b}^{\dagger}_n }
  \frac{1}{\sqrt{2}} \left( | 0 \rangle_T + | \pi r \rangle_T \right),
\\
| N(\frac{\pi}{2 r})_T \rangle &=&
  e^{ + \sum_{n=1}^{\infty} b^{\dagger}_n \tilde{b}^{\dagger}_n }
  \frac{1}{\sqrt{2}} \left( | 0 \rangle_T - | \pi r \rangle_T \right).
\end{eqnarray}

While these four twisted boundary states satisfy the Ishibashi condition,
they are not consistent boundary states by themselves.
The amplitudes among them are
\begin{eqnarray}
 \langle D( \varphi_0 )_T | e^{ - H^{(A)}_{\beta}/2} | D(\varphi_0)_T \rangle
&=& \tilde{q}^{1/48} \prod_{n=1}^{\infty} \frac{1}{1-\tilde{q}^{n-1/2}}
\nonumber \\
&=& \frac{\vartheta_2(\tilde{q}^{1/2})}{2 \eta (\tilde{1}) }
\nonumber \\
&=& \frac{\vartheta_4(q^2)}{\sqrt{2} \eta (q)} 
\label{eq:DTDTamp}
\\
&=& 
 \langle N (\tilde{\varphi}_0)_T | e^{ - H^{(A)}_{\beta}/2} 
        | N (\tilde{\varphi}_0)_T \rangle
\nonumber \\
 \langle D( \varphi_0 )_T | e^{ - H^{(A)}_{\beta}/2} 
          | N(\tilde{\varphi}_0)_T \rangle
&=& \pm \frac{1}{\sqrt{2}} \tilde{q}^{1/48}
    \prod_{n=1}^{\infty} \frac{1}{1 + \tilde{q}^{n-1/2}}
\nonumber \\
&=& \pm \frac{1}{2 \eta (\tilde{q})} \vartheta_2(e^{i \pi/2},\tilde{q}^{1/2})
\nonumber \\
&=& \pm \frac{1}{ \sqrt{2} \eta (q)} q^{1/16} \vartheta_4(q^{-1/2},q^2),
\label{eq:DTNTamp}
\end{eqnarray}
where $\varphi_0 = 0$ or $\pi r$, $\tilde{\varphi}_0 = 0$ or $\pi/(2r)$
and $\pm$ depends on the value of $\varphi_0$ and $\tilde{\varphi}_0$.
The other amplitudes vanish:
\begin{eqnarray}
 \langle D( 0 )_T | e^{ - H^{(A)}_{\beta}/2} | D(\pi r)_T \rangle
&=& 
 \langle N ( 0 )_T | e^{ - H^{(A)}_{\beta}/2} 
        | N ( \frac{\pi}{2 r} )_T \rangle
\nonumber \\
&=& 0 .
\end{eqnarray}

In order to construct consistent boundary states,
we must combine the twisted and untwisted sectors
with appropriate coefficients.
These endpoint Dirichlet/Neumann boundary states are 
\begin{eqnarray}
| D_O (\varphi_0) \pm \rangle & \equiv & 
  2^{-1/2} | D( \varphi_0 ) \rangle \pm 2^{-1/4} | D( \varphi_0 )_T \rangle ,
\label{eq:discDBS}
\\
| N_O (\tilde{\varphi}_0) \pm \rangle & \equiv & 
  2^{-1/2} | N( \tilde{\varphi}_0 ) \rangle \pm 2^{-1/4}
                           | N( \tilde{\varphi}_0 )_T \rangle ,
\label{eq:discNBS}
\end{eqnarray}
where $\varphi_0$ and $\tilde{\varphi}_0$ again take only the endpoint
values.
The amplitude among those endpoint Dirichlet boundary states are
\begin{eqnarray}
\langle D_O (\varphi_0) \pm | e^{ - H^{(A)}_{\beta}/2} 
        | D_O (\varphi_0) \pm  \rangle
&=& 
\frac{1}{2} Z_r(0) + \frac{1}{2 \eta(q)} \vartheta_4 (q^2)
\nonumber \\
&=& \frac{1}{2 \eta(q)} \sum_{n = - \infty}^{\infty}
    \left[   q^{2 r^2 n^2} + (-1)^n q^{n^2} \right]
\nonumber \\
&=&  \sum_{n = 1}^{\infty} \chi_{2 r^2 n^2}
    + \sum_{n=0}^{\infty} \chi_{(2n)^2},
\\
\langle D_O (\varphi_0) \pm | e^{ - H^{(A)}_{\beta}/2} 
        | D_O (\varphi_0) \mp  \rangle
&=&
\frac{1}{2} Z_r(0) - \frac{1}{2 \eta(q)} \vartheta_4 (q^2)
\nonumber \\
&=& \frac{1}{2 \eta(q)} \sum_{n = - \infty}^{\infty}
    \left[   q^{2 r^2 n^2} - (-1)^n q^{n^2} \right]
\nonumber \\
&=&  \sum_{n = 1}^{\infty} \chi_{2 r^2 n^2}
    + \sum_{n=0}^{\infty} \chi_{(2n+1)^2},
\\
\langle D_O (\varphi_0) \nu_1 | e^{ - H^{(A)}_{\beta}/2} 
        | D_O (\pi r - \varphi_0) \nu_2  \rangle
&=&
\frac{1}{2} Z_r (\pi r)
\nonumber \\
&=& \sum_{n=1}^{\infty} \frac{q^{r^2(2n-1)^2}}{\eta (q)},
\end{eqnarray}
where $\nu_{1,2}$ take either $+$ or $-$ independently.
The amplitude among the endpoint Neumann are given by replacing
the radius $r$ by its dual $1/(2r)$.
The amplitudes between the endpoint Dirichlet and Neumann
boundary states are given by
\begin{eqnarray}
\langle D_O (\varphi_0) \nu_1 | e^{ - H^{(A)}_{\beta}/2} 
        | N_O (\tilde{\varphi}_0) \nu_2  \rangle
&=& 
\frac{1}{4 \eta (q)} \vartheta_2 (q)
\pm \frac{1}{ 2 \eta (q)} q^{1/16} \vartheta_4(q^{-1/2},q^2)
\nonumber \\
&=&  \sum_{n=0}^{\infty} \chi_{\frac{1}{2} (2n -1 \pm \frac{1}{2})^2 } ,
\end{eqnarray}
where $\pm$ in the last line depends on the signs $\nu_{1,2}$ and values
of $\varphi_0$ and $\tilde{\varphi}_0$.
Thus these eight discrete boundary states  mutually consistent.
Moreover, they are consistent with the continuous
Dirichlet~(\ref{eq:bstateboson})
and von Neumann~(\ref{eq:orbNeuBS}) boundary states on the orbifold.
This can be easily seen because only the untwisted sector
in~(\ref{eq:discDBS}) and~(\ref{eq:discNBS})
contribute to the mutual amplitude with the continuous ones.
For example,
\begin{equation}
\langle D_O (0) \pm | e^{ - H^{(A)}_{\beta}/2} 
        | D_O (\varphi_0) \rangle = Z_r(\varphi_0) .
\end{equation}

In this way, we have found two continous families of boundary states
and eight discrete ones and they are all mutually consistent boundary
states.
While our construction does not exclude the presence of other boundary
states, we conjecture that the above exhaust all the possible
orbifold boundary states for generic values of the compactification
radius $r$.
An analogous statement has been proven for a non-orbifold
boson~\cite{Friedan}.
(For special values of $r$, another family of boundary
states is constructed for non-orbifold
boson~\cite{Callan:SU2boundary}.
Similar construction may apply to the orbifold boson.)

Finally, we note that although the above orbifold boundary states
are mutually complatible,
they are incompatible with non-orbifold boson boundary states.
For example, the coefficient of
the boson Dirichlet boundary state is fixed as~(\ref{eq:DirBS}),
by requiring the compatibility with itself.
Thus the partition function with the boundary conditions
(\ref{eq:bstateboson}) and (\ref{eq:DirBS}) at the opposite sides
is given by
 \begin{equation}
\frac{1}{\sqrt{2}} 
  \left[ Z_r(\varphi_0 - \varphi'_0) + Z_r(\varphi_0 + \varphi'_0) 
  \right] .
\end{equation}
This contains non-integer (irrational) coefficient when expressed
as a sum of $c=1$ Virasoro characters;
the two boundary conditions are incompatible.
This incompatibility is presumably related to the fact that
the sets of local operators are different for the free boson theory
and its orbifold (Ashkin-Teller model).

\subsection{Applications to the defect line problem} 

Now we turn to our original problem of the Ising model with
a defect line.
As explained, the bulk theory in the present problem is the
special $r=1$ point of the orbifold, after the folding.
Thus a universality class of the defect line should correspond to
an orbifold boundary state with $r=1$.
Hereafter we fix the compactification radius as $r=1$.

From the general discussion of the orbifold boundary states,
we have two continuous family of boundary states and
eight discrete ones, all of which are mutually consistent.
The continuous Dirichlet boundary state~(\ref{eq:bstateboson})
is identified with the boundary state~(\ref{eq:bstateAT})
for the defect line in~(\ref{eq:defectham}),
using the same $\varphi_0$ defined in~(\ref{eq:phi0def}).
In fact, the partition functions constructed from~(\ref{eq:bstateAT})
and~(\ref{eq:bstateboson}) agrees completely, observing that
$Z(\Delta \varphi_0)$ in eq.~(\ref{eq:Zofphi}) is identical
to $Z_r(\Delta \varphi_0)$ in eq.~(\ref{eq:Zrboson}) for $r=1$.

Comparing~(\ref{eq:bstateAT}) and~(\ref{eq:bstateboson}), we can
identify the symmetric combination of bosonic oscillator vacua as
\begin{equation}
\frac{1}{\sqrt{2}} \left[ | (0,k) \rangle + | (0, -k) \rangle \right]
= | \frac{k^2}{8} , S \rangle,
\end{equation}
where the right hand side is defined as eq.~(\ref{eq:bstateSym})
in the Ashkin-Teller classification.
The other symmetric combination is also identified with eq.~(\ref{eq:ATBSA}) as
\begin{equation}
\frac{1}{\sqrt{2}} \left[ | (n,0) \rangle + | (-n, 0) \rangle \right]
= | \frac{n^2}{2} , A \rangle.
\end{equation}

Other $3 \times 3 =9$ universality classes can be readily
obtained by cutting the system at the defect
line, and imposing one of the three possible Ising boundary universality
class on the two sides.
As discussed in the previous section,
the corresponding boundary states are given by the tensor product
of the Ising boundary states.
One of them, $| ff \rangle$ is actually the special case
$\varphi_0 = \pi/2$ of the Dirichlet boundary state~(\ref{eq:bstateAT})
or~(\ref{eq:bstateboson}).
The remaining 8 universality classes nicely fit with the 8 discrete
endpoint boundary states of the orbifolds.
Actually we can identify them as following:
\begin{eqnarray}
| \uparrow \uparrow \rangle &=& | D_O (0) + \rangle
\\
| \uparrow \downarrow \rangle &=&  | D_O (\pi) - \rangle
\\
| \downarrow \uparrow \rangle &=&  | D_O (\pi) + \rangle
\\
| \downarrow \downarrow \rangle &=& | D_O (0) - \rangle 
\\
| \uparrow f \rangle &=&  | N_O (0) + \rangle
\label{eq:upforb}
\\
| \downarrow f \rangle &=& | N_O (0) - \rangle
\\
| f \uparrow \rangle &=& | N_O (\frac{\pi}{2}) + \rangle
\\
| f \downarrow \rangle &=& | N_O (\frac{\pi}{2}) - \rangle 
\label{eq:fdnorb}
\end{eqnarray}
We note that the sign flip of the coefficient of the
twisted sector corresponds to an overall Ising spin flip.

These are equivalent to the following identification
between the twisted orbifold boundary state
and the Ashkin-Teller Ishibashi states:
\begin{eqnarray}
| D(0)_T  \rangle 
&=& \frac{1}{\sqrt{2}} \sum_{n=0}^{\infty} 
       \left(   | \frac{(2n+1)^2}{16}, 1 \rangle \rangle 
              + | \frac{(2n+1)^2}{16}, 2 \rangle \rangle 
       \right),
\\
| D(\pi)_T  \rangle 
&=& \frac{1}{\sqrt{2}} \sum_{n=0}^{\infty} 
       \left(   | \frac{(2n+1)^2}{16}, 1 \rangle \rangle 
              - | \frac{(2n+1)^2}{16}, 2 \rangle \rangle 
       \right),
\\
| N(0)_T  \rangle 
&=& \sum_{n=0}^{\infty} (-1)^{n(n+1)/2}
    | \frac{(2n+1)^2}{16}, 1 \rangle \rangle 
\\
| N(\frac{\pi}{2})_T  \rangle 
&=& \sum_{n=0}^{\infty} (-1)^{n(n+1)/2}
    | \frac{(2n+1)^2}{16}, 2 \rangle \rangle 
\end{eqnarray}
These correspondences are verified from the calculation of the
partition functions:
\begin{eqnarray}
\sum_{n=0}^{\infty} \langle \langle \frac{(2n+1)^2}{16} |
 \tilde{q}^{(L_0 + \bar{L}_0)/2} | \frac{(2n+1)^2}{16} \rangle \rangle
&=& \frac{1}{2 \eta (\tilde{q})} \vartheta_2 (\tilde{q}^{1/2}),
\\
\frac{1}{\sqrt{2}}
\sum_{n=0}^{\infty} \langle \langle \frac{(2n+1)^2}{16} |
 (-1)^{\frac{n(n+1)}{2}}
 \tilde{q}^{\frac{(L_0 + \bar{L}_0)}{2}} 
 | \frac{(2n+1)^2}{16} \rangle \rangle
&=& \frac{1}{2 \eta(\tilde{q})}
       \vartheta_2 (e^{\pi i/2}, \tilde{q}^{1/2}),
\end{eqnarray}
which should be compared with~(\ref{eq:DTDTamp}) and~(\ref{eq:DTNTamp}).

Thus all the known defect universality classes are understood
in terms of orbifold boundary states.
On the other hand, the defect line corresponding to
the continuous von Neumann boundary
state~(\ref{eq:orbNeuBS}) has not been yet identified.
This actually gives a previously unknown universality class
of defect lines.
The nature of this defect universality class will be discussed
in Section~\ref{sec:Ncorr}.

\section{Correlation functions for the continuous Dirichlet universality class}
\label{sec:Dcorr}

Here we discuss several correlation functions in the presence
of the defect line~(\ref{eq:defectham}) that is identified
with the continuous Dirichlet boundary state~(\ref{eq:bstateboson})
or~(\ref{eq:bstateAT}) in our approach.

\subsection{Spectrum of boundary operators}
\label{subsec:spectrum}

As discussed in Section~\ref{sec:bstateorb},
the complete spectrum of boundary operators
can be determined from the partition function
of the strip with the same boundary condition
at the both sides~\cite{Cardy:fusion}.
In the case of the continuous Dirichlet universality
class~(\ref{eq:bstateboson}),
it is obtained by putting $\varphi'_0 = \varphi_0$ and $r=1$ in
eq.~(\ref{eq:Zphiphi}) as
\begin{equation}
  Z = Z(0) + Z( 2 \varphi_0 ),
\end{equation}
where $Z(\Delta \varphi_0)$ is defined in~(\ref{eq:Zofphi}).
By expressing this result in terms of
$c=1$ Virasoro characters~(\ref{eq:c1character}),
we obtain the complete set of
boundary primary operators.

The boundary operators consist of two sets:
the one corresponds to $Z(0)$, which is independent of the boundary
condition (defect strength),
and the other corresponds to
$Z(2 \varphi_0)$ which depends on the boundary condition.
As we discuss later,
the former is identified with the boundary operators associated with
the bosonic operators, and the latter is the boundary operators
associated with twist (spin) operators.
The surface critical exponent of the operator is defined by
the correlation function of the two operators located
far apart along the boundary (the defect line in our case). 
This surface critical exponent is given by the dimension of the
corresponding boundary operator.
Here we give the complete spectrum of the boundary primary operators
in the ``moving sector'' $Z(2 \varphi_0)$:
\begin{equation}
  \Delta_b = 2n^2 + \frac{4 n \varphi_0}{\pi} + 2 \frac{{\varphi_0}^2}{\pi^2},
\label{eq:movingsector}
\end{equation}
where $n$ is an arbitrary integer.
(The above is the complete set for generic $\varphi_0$, where no
degenerate representation of $c=1$ Virasoro algebra occurs.)

Let us assume that the surface exponent of the Ising spin is given by
the boundary operator with the smallest dimension in the above
``moving'' sector.
Then we find that the exponent depends continuously on the defect strength as
\begin{equation}
2 \Delta_b = \mbox{min} \left(
   4 (\frac{\varphi_0}{\pi})^2 , 4 (\frac{\pi - \varphi_0}{\pi})^2 
   \right) .
\label{eq:boundaryexp}
\end{equation}
This continuously changing surface exponent
reproduces the previous result by Bariev~\cite{Bariev:Ising} and
McCoy and Perk~\cite{McCoyPerk}.
(Note that their lattice model is dual to our model, and
surface exponent changes under the duality transformation
as discussed in Ref.~\cite{Brown:defect}.
In Section~\ref{subsec:dual}, we 
argue they belong to the same universality class with
different parameters.)

Our formalism gives a complete set of boundary operators
in addition to their result.
Note that the surface exponent is unchanged under
$\varphi_0 \rightarrow \pi - \varphi_0$ which corresponds to
the sign reversal of the defect $b \rightarrow -b$.

We note that, for $\varphi_0 = 0$ or $\pi$ there is a dimension zero
operator other than the identity.
Thus Cardy's consistency condition does not hold in these cases.
The presence of the dimension zero operator is related to the
infinitely long range correlation of the spin operators along the
boundary.
This somewhat unphysical situation
is only achieved as a limit of anisotropic and strong defect limit.
(However, it corresponds to $b=0$ in dual quantum model~(\ref{eq:dualqham}),
which is a well-defined quantum model.)

\subsection{Two-spin correlation functions}
\label{subsec:spincorr}

The conformal field techniques enable us to calculate
also the exact two-point correlation functions at general location
for arbitrary strength of the defect.
Let us consider the two-spin correlation function.
By folding, the correlation of two spins located on the same or opposite
side of the boundary is mapped to the correlation
function $\langle \sigma_1 \sigma_1 \rangle$
or $\langle \sigma_1 \sigma_2 \rangle$, respectively.
Let us take the boundary as the real axis of the complex plane.
In boundary conformal field theory, a basic technique is to
analytically continue the energy-momentum tensor beyond the boundary.
Then the two-point correlation function
of a non-chiral operator is mapped
to a four-point correlation function of chiral operators as
\begin{equation}
  \langle \sigma_j (z_1,\bar{z}_1) \sigma_k (z_2,\bar{z}_2) \rangle
=
  \langle \sigma_j (z_1)  \sigma'_j (z^*_1)
              \sigma_k(z_2) \sigma'_k(z^*_2) \rangle,
\label{eq:mirrorcorr}
\end{equation}
where $z^*_1$ and $z^*_2$ are complex conjugates of $z_1$ and $z_2$,
and $\sigma_j(z_l)$ is a chiral operator which should be distinguished
from the non-chiral $\sigma_j(z_l,\bar{z}_l)$.
This may be regarded as a generalization of the mirror image method.

In the present problem, the non-chiral spin operator
$\sigma_j(z,\bar{z})$ has the
dimension $(1/16,1/16)$ and all the four operators
in~(\ref{eq:mirrorcorr}) have the dimension $1/16$. 
Thus the correlation function should be constructed from
the $c=1$ conformal blocks of four spin operators of dimension $1/16$.
These conformal blocks are obtained by Zamolodchikov~\cite{Zam:ATspin} as
\begin{equation}
  f(\Delta,\frac{1}{16},1,x) =
   (16 \sqrt{u})^{\Delta} \vartheta_3^{-1}(u)
\end{equation}
where $\Delta$ is the dimension of the intermediate primary state,
$x$ is the cross-ratio
\begin{equation}
  x = \left| \frac{z_1-z_2}{z_1 - z_2^*} \right| ^2
\label{eq:crossratio}
\end{equation}
and $u$ is determined from $x$ by
\begin{equation}
  x = \left( \frac{\vartheta_2(u)}{\vartheta_3(u)} \right) ^4 .
\label{eq:udef} 
\end{equation}
(Note that our definition of the theta functions is different from
that used in Ref.~\cite{Zam:ATspin}.)
Any real dimension $\Delta$ of the intermediate state
is allowed in this $c=1$ theory.

The correlation function can be determined if the intermediate 
states and the corresponding OPE coefficients are identified.
These OPE coefficients depend on the boundary condition (defect strength)
and in general different from those in the bulk four-point
function.
Nevertheless, Cardy and Lewellen~\cite{CardyLewellen} presented a
method to determine them from the bulk four-point function and
the boundary state.
We use this idea to determine
the boundary correlation function~(\ref{eq:mirrorcorr}).

The essential point of their argument is
that the OPE of the two bulk operators is independent
of the boundary condition.
If we use the bulk OPE, the boundary
correlation function is in principle
obtained from the one-point function in the presence
of the boundary:
\begin{equation}
  \langle \sigma_j(z_1,\bar{z}_1) \sigma_k(z_2,\bar{z}_2)
            \rangle \sim
\sum_{\Delta,\bar{\Delta}} C_{\sigma_j \sigma_k (\Delta, \bar{\Delta})}
      \langle \phi_{\Delta, \bar{\Delta}} + \mbox{(descendants)}
             \rangle ,
\end{equation}
where $\phi_{\Delta, \bar{\Delta}}$ is a primary field with dimension
$(\Delta, \bar{\Delta})$, $C_{\sigma_j \sigma_k (\Delta, \bar{\Delta})}$
are the bulk OPE coefficients
and the coordinate dependent factor is omitted.
The one-point boundary correlation function of the primary field
$\phi_{\Delta, \bar{\Delta}}$ is given in terms of the
boundary state as
\begin{eqnarray}
 \langle \phi_{\Delta, \bar{\Delta}} (y) \rangle &=&
      A^{\phi}_B  \left( \frac{1}{2y} \right)^{\Delta + \bar{\Delta}} \nonumber
\\
&=&
   \frac{\langle \phi_{\Delta \bar{\Delta}} | B \rangle}{\langle 0 | B \rangle}
            \left( \frac{1}{2y} \right) ^{\Delta + \bar{\Delta}},
\label{eq:boundary1pt}
\end{eqnarray}
where $y$ is the distance from the boundary.
The boundary state $| B \rangle$ consists of Ishibashi states.
In general, only the primary states $| \phi_{\Delta \bar{\Delta}} \rangle$
with $\Delta = \bar{\Delta}$ can have non-vanishing $A^{\phi}_B$. 
Once the contribution from an intermediate primary state is identified
in this way, contributions from its descendants are determined
from the conformal invariance.
The total contribution from a conformal family should be
the conformal block function multiplied by the constant $A^{\phi}_B$,
as required from eq.~(\ref{eq:mirrorcorr}).

Zamolodchikov~\cite{Zam:ATspin} also
constructed an explicit expression of the bulk correlation function
of four spin operators of the Ashkin-Teller model in the same layer.
In the doubled Ising case, it reads
\begin{eqnarray}
\lefteqn{  \langle \sigma_1 (z_1, \bar{z}_1) \sigma_1 (z_2,\bar{z}_2)
          \sigma_1 (z_3, \bar{z}_3) \sigma_1 (z_4,\bar{z}_4) \rangle =}
\nonumber \\
&& | \frac{z_{14} z_{32} }{z_{12} z_{34} z_{13} z_{42}} |^{1/4}
\frac{1}{\vartheta_3(u)}
\sum_{k,l = -\infty}^{\infty}
     [ \sqrt{u}^{2k^2} \sqrt{\bar{u}}^{2l^2} +
       \sqrt{u}^{(2k+1)^2/2} \sqrt{\bar{u}}^{(2l+1)^2/2} ],
\label{eq:bulk11}
\end{eqnarray}
which is equivalent to the well-known four-point
function of the Ising model.
From this correlation function, we can read off
the dimensions of the intermediate primaries as
\begin{equation}
  (\Delta, \bar{\Delta}) = (2k^2 , 2l^2) \;\;\;\;\;
                 (\frac{(2k+1)^2}{2},\frac{(2l+1)^2}{2} ) .
\end{equation}
In the calculation of the boundary correlation functions,
we need only the spinless primaries
\begin{equation}
  \Delta = \bar{\Delta} = \frac{k^2}{2}
\label{eq:dimOPE11}
\end{equation}
and the corresponding OPE coefficient.

To calculate the correlation function of the spins in the opposite side
of the defect line ($\langle \sigma_1 \sigma_2 \rangle$
after the folding),  we need the OPE of $\sigma_1$ and $\sigma_2$.
This can be read off from the bulk four-point function
$\langle \sigma_1 \sigma_2 \sigma_1 \sigma_2 \rangle$.
Since the two layers are not coupled in the bulk,
this is simply a product of two-point functions 
$\langle \sigma_1 \sigma_1 \rangle \langle \sigma_2 \sigma_2 \rangle$.
Still it can also be written in terms of the $c=1$ conformal block as
\begin{eqnarray}
\lefteqn{  \langle \sigma_1 (z_1, \bar{z}_1) \sigma_2 (z_2,\bar{z}_2)
          \sigma_1 (z_3, \bar{z}_3) \sigma_2 (z_4,\bar{z}_4) \rangle =}
\nonumber \\
&& \left| \frac{z_{14} z_{32} }{z_{12} z_{34} z_{13} z_{42}} \right|^{1/4}
\frac{2}{\vartheta_3(u)}
\sum_{k,l = 1}^{\infty} 
 \sqrt{u}^{(2k-1)^2/8} \sqrt{\bar{u}}^{(2l-1)^2/8}  .
\label{eq:bulk12}
\end{eqnarray}
Thus the spinless primaries appear in the OPE of $\sigma_1 \sigma_2$ are
\begin{equation}
  \Delta = \bar{\Delta} = \frac{(2k-1)^2}{8} .
\label{eq:dimOPE12}
\end{equation}

There is some ambiguity in determination of the
OPE coefficient from the bulk correlation functions.
The sign of the OPE coefficient cannot be determined from the bulk
correlation where the square of a coefficient appears.
The sign actually depends on the definition of the intermediate
field,  but the boundary correlation function depends only on the product
$A^{\phi}_B C_{\sigma \sigma \phi}$, which is independent of the
sign convention of the field $\phi$.
In the following, we fix the sign convention so that $A^{\phi}_B \geq 0$.
Anyway we cannot determine the sign of $A^{\phi}_B C_{\sigma \sigma \phi}$
from those bulk correlation functions.

Thus we assume a specific form of the OPE coefficients
that is consistent with the bulk correlation functions,
and construct the boundary two-point function.
Then we will check the validity of our assumption
by comparing the result with the known cases.

We assume all the OPE coefficients are positive.
The spinless operators appearing in the OPE of $\sigma_1$ and $\sigma_2$
have dimensions as shown in eq.~(\ref{eq:dimOPE12}) and are identified with the
third set of the Ashkin-Teller primaries listed in Table~\ref{tab:ATops}.
The bulk correlation function~(\ref{eq:bulk12}) uniquely determines
the OPE coefficients as
\begin{equation}
C_{\sigma_1 \sigma_2 (2n-1)^2/8}
   =  \sqrt{2} \left( \frac{1}{16} \right)^{(2n-1)^2/8} 
\end{equation}
where $(2n-1)^2/8$ is the primary weight of the intermediate operator.
The spinless operators in the OPE of $\sigma_j$ and $\sigma_j$
($j=1$ or $2$)
have dimensions~(\ref{eq:dimOPE11}).
They are naturally identified with $n=0$ in the first set
and the whole second set.
This is consistent with the Ising OPE: $ \sigma \sigma \sim [I] + [\epsilon]$.
The second set with the primary weight $n^2/2$
has multiplicity two for each dimension, as we discussed
in the last section.
We determine which operator appears in the OPE, from
the Ising OPE.
For odd $n = 2k+1$ (weight $(2k+1)^2/2$),
they are identified as the descendants of
$\epsilon_1 \otimes I_2$ and $I_1 \otimes \epsilon_2$.
Thus the OPE of $\sigma_1$ and $\sigma_1$ produces the former,
and that of $\sigma_2$ and $\sigma_2$ produces the latter.
For even $n = 2k$ (weight $2k^2$),
they are identified as the descendants of
$I_1 \otimes I_2$ and $\epsilon_1 \otimes \epsilon_2$.
From the Ising OPE, we see that both OPE of $\sigma_j$ and $\sigma_j$
($j=1,2$) produces the former.
Since the boundary state~(\ref{eq:bstateAT}) consists only the
symmetric combination of corresponding Ishibashi
states as in~(\ref{eq:bstateSym}),
we consider the OPE coefficient for the symmetrized operator.
It is determined from the bulk correlation
function~(\ref{eq:bulk11}) as
\begin{equation}
C_{\sigma_j \sigma_j (n^2/2, S)}
 =  
\sqrt{2} \left( \frac{1}{16} \right)^{n^2/2}
\end{equation}
for $j=1,2$ and $n \geq 1$.
The coefficient for identity operator is
\begin{equation}
  C_{\sigma_j \sigma_j 0} = 1
\end{equation}
for $j=1,2$.

The coefficients $A^{\Delta}_B$ is read off from (\ref{eq:bstateAT}) as
\begin{equation}
  A^{(n^2/8, S)}_B = \left\{  \begin{array}{ll}
                              1 & (n=0) \\
                      \sqrt{2} \cos{(n \varphi_0)} & (n \geq 1) 
                            \end{array} \right.  
\end{equation}
Therefore we obtain the boundary correlation functions
as
\begin{eqnarray}
\lefteqn{  \langle \sigma_j (z_1,\bar{z}_1) \sigma_j (z_2,\bar{z}_2) \rangle }
\nonumber \\
&=& \left( \frac{1}{4 y_1 y_2 x} \right)^{1/8} \frac{1}{\vartheta_3(u)}
   \sum_{n=0}^{\infty} A^{(n^2/2,S)}_B C_{\sigma_j \sigma_j (n^2/2,S)}
                           f(\frac{n^2}{2},\frac{1}{16},1,x)
\nonumber \\
&=& 
\left( \frac{1}{4 y_1 y_2 x} \right)^{1/8} \frac{1}{\vartheta_3(u)}
   ( 1 + \sum_{n=1}^{\infty} 2 \cos{(2 n \varphi_0)} u^{\frac{n^2}{4}} )
\nonumber \\
&=&
\left( \frac{1}{4 y_1 y_2 x} \right)^{1/8} \frac{1}{\vartheta_3(u)}
          \vartheta_3(e^{2 i \varphi_0}, \sqrt{u}), 
\label{eq:boundary11}
\\
\lefteqn{ \langle \sigma_1 (z_1,\bar{z}_1) \sigma_2 (z_2,\bar{z}_2) \rangle}
\nonumber \\
&=&
\left( \frac{1}{4 y_1 y_2 x} \right)^{1/8} \frac{1}{\vartheta_3(u)}
       \sum_{n=0}^{\infty} A^{(2n+1)^2/8}_B C_{\sigma_1 \sigma_2 (2n+1)^2/8}
                           f(\frac{(2n+1)^2}{8},\frac{1}{16},1,x)
\nonumber \\
&=& 
\left( \frac{1}{4 y_1 y_2 x} \right)^{1/8} \frac{1}{\vartheta_3(u)}
          \vartheta_2(e^{2 i \varphi_0}, \sqrt{u}),
\label{eq:boundary12}
\end{eqnarray}
where $y_1$ and $y_2$ are the imaginary parts of $z_1$ and $z_2$,
$x$ is the cross-ratio~(\ref{eq:crossratio}),
and $u$ is a function of $x$ determined by eq.~(\ref{eq:udef}).
This generalizes Brown's result~\cite{Brown:defect}, which is the first
order perturbation in the defect strength.

Let us check the result, in order to justify our assumptions about the
OPE coefficients.
\subsubsection*{No defect (b=1)}
\begin{eqnarray}
  \langle \sigma_1 \sigma_1 \rangle &=&
       \frac{1}{|z_1 - z_2|^{1/4}} \frac{1}{(1-x)^{1/8}}   
       \frac{1}{\vartheta_3(u)} \vartheta_3(e^{i \pi /2},\sqrt{u})
\nonumber \\
&=&    \frac{1}{|z_1 - z_2|^{1/4}} 
       \frac{1}{\sqrt{ \vartheta_3(u) \vartheta_4(u)}} \vartheta_4(u^2)
\nonumber \\
&=&   \frac{1}{|z_1 - z_2|^{1/4}}    
\\
\langle \sigma_1 \sigma_2 \rangle &=&
       \frac{1}{|z_1 - z^*_2|^{1/4}} \left[ \frac{1}{x(1-x)} \right]^{1/8}   
       \frac{1}{\vartheta_3(u)} \vartheta_2(e^{i \pi /2},\sqrt{u})
\nonumber \\
&=&    \frac{1}{|z_1 - z^*_2|^{1/4}} 
       \frac{1}{\sqrt{ \vartheta_2(u) \vartheta_4(u)}} \vartheta_2(e^{i \pi /4},\sqrt{u})
\nonumber \\
&=&   \frac{1}{|z_1 - z^*_2|^{1/4}}    
\end{eqnarray}
Thus we recover the known result (bulk two-spin correlation)
for both $\langle \sigma_1 \sigma_1 \rangle$ and
$\langle \sigma_1 \sigma_2 \rangle$.

\subsubsection*{Free boundary ($b=0$)}
\begin{eqnarray}
  \langle \sigma_1 \sigma_1 \rangle &=&
       \frac{1}{(4 y_1 y_2)^{1/8}} \frac{1}{x^{1/8}}
       \frac{1}{\vartheta_3(u)} \vartheta_3(e^{i \pi},\sqrt{u})
\nonumber \\
&=&
       \frac{1}{(4 y_1 y_2)^{1/8}}
       \frac{1}{\sqrt{ \vartheta_2(u) \vartheta_3(u)}} \vartheta_4(\sqrt{u})
\nonumber \\
&=&
       \frac{1}{(4 y_1 y_2)^{1/8}}  \sqrt{x^{-1/4} - x^{1/4}}
\\
\langle \sigma_1 \sigma_2 \rangle &=&
       \frac{1}{|z_1 - z^*_2|^{1/4}} 
       \left[ \frac{1}{x(1-x)} \right]^{1/8}   
       \frac{1}{\vartheta_3(u)} \vartheta_2(e^{i \pi },\sqrt{u})
\nonumber \\
&=& 0 
\end{eqnarray}
The former agrees with the boundary two-point correlation of the
Ising model with the free boundary condition~\cite{Cardy:BCFT}.
The latter result is expected since the two layers are completely
decoupled at this point.

\subsubsection*{Antiferromagnetic defect ($b=-1$)}
\begin{eqnarray}
  \langle \sigma_1 \sigma_1 \rangle &=&
       \frac{1}{|z_1 - z_2|^{1/4}} \frac{1}{(1-x)^{1/8}}   
       \frac{1}{\vartheta_3(u)} \vartheta_3(e^{i 3 \pi /2},\sqrt{u})
\nonumber \\
&=&    \frac{1}{|z_1 - z_2|^{1/4}} 
       \frac{1}{\sqrt{ \vartheta_3(u) \vartheta_4(u)}} \vartheta_4(u^2)
\nonumber \\
&=&   \frac{1}{|z_1 - z_2|^{1/4}}    
\\
\langle \sigma_1 \sigma_2 \rangle &=&
       \frac{1}{|z_1 - z^*_2|^{1/4}} \left[ \frac{1}{x(1-x)} \right] ^{1/8}   
       \frac{1}{\vartheta_3(u)} \vartheta_2(e^{i 3\pi /2},\sqrt{u})
\nonumber \\
&=&  -  \frac{1}{|z_1 - z^*_2|^{1/4}} 
  \frac{1}{\sqrt{ \vartheta_2(u) \vartheta_4(u)}} \vartheta_2(e^{i \pi /2},\sqrt{u})
\nonumber \\
&=&  -  \frac{1}{|z_1 - z^*_2|^{1/4}}    
\end{eqnarray}
The former is the same as the no defect case ($b=0$), while the latter
changes sign.
This is consistent with the original lattice model,
where one can reverse the sign of all spins
on one side of the defect to change the sign of the defect strength.
Thus our formulae again give the correct description of the system.

\begin{figure}[htbp]
  \begin{center}
    \leavevmode
    \epsfxsize=\textwidth
    \epsfbox{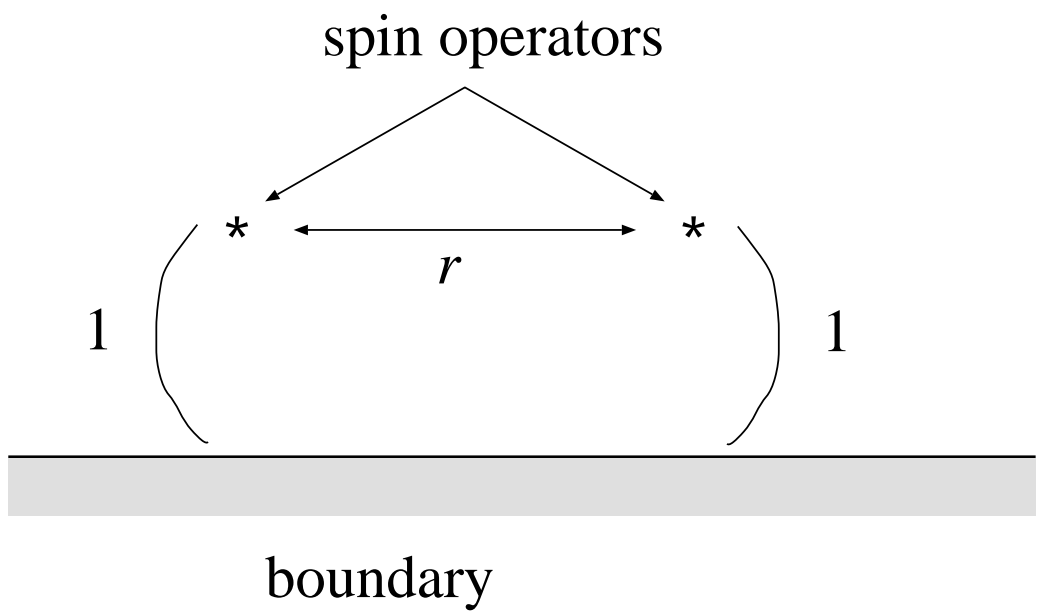}
\caption{
The location of spin operators used in the graph of correlation
functions (Fig.~\protect\ref{fig:graph}).
We fix
the distance of spin operators from the boundary to $1$,
and leave the horizontal distance $r$ between them as a variable.
}
    \label{fig:spinloc}
  \end{center}
\end{figure}

\begin{figure}[htbp]
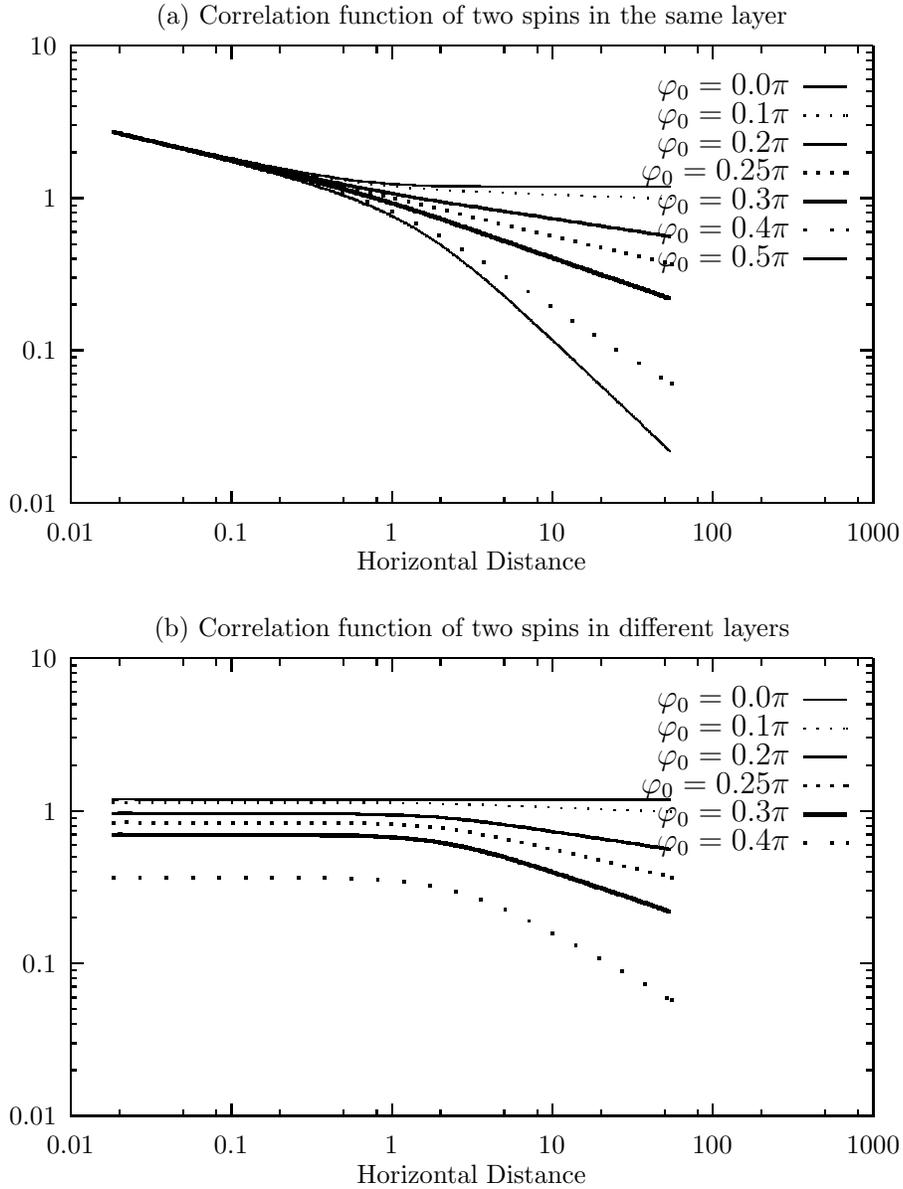

  \begin{center}
    \leavevmode
\input{c11.tex}
    \vspace*{0.5cm}
\input{c12.tex}
\caption{The two-spin correlation for various strength of the defect.
We show the result on (a) $\langle \sigma_1 \sigma_1 \rangle$ and
(b) $\langle \sigma_1 \sigma_2 \rangle$ for
$\varphi_0 = 0$ (strong coupling and anisotropic limit), 
$0.1 \pi , 0.2 \pi , 0.25 \pi$ (no defect)  , $0.3 \pi , 0.4 \pi$
and $0.5 \pi$ (free boundary condition).
They are shown as a function of the
distance $r$, in a log-log plot.}
  \label{fig:graph}
  \end{center}
\end{figure}

From the agreement with the known cases, we believe
our exact expressions~(\ref{eq:boundary11}) and (\ref{eq:boundary12})
of the boundary two-spin correlation are correct, though we made some
assumptions during the calculation.
While the parameter $u$ is implicitly defined
by the eq.~(\ref{eq:udef}), there is no essential difficulty
in solving the equation numerically.
We used Mathematica to solve the equation
and then draw graphs of our results.
We choose the location of the spins as shown in Fig.~\ref{fig:spinloc}. 
We show the correlation functions
for various strength of the defect in Fig.~\ref{fig:graph}
as a function of the distance.
In the short-distance limit,
$\langle \sigma_1 \sigma_1 \rangle$ converges to a single power-law
which is independent of the defect strength.
This is the expected bulk limit.
In the large-distance limit, the correlation function is governed
by another exponent, which depends on the defect strength.
This is the boundary limit.
In the next subsection, we will find the boundary exponent
actually agrees with the boundary operator with the smallest dimension
found in eq.~(\ref{eq:boundaryexp}).
Our result interpolates between these two limits.

In the ``short-distance'' limit,
$\langle \sigma_1 \sigma_2 \rangle$ converges to a constant
which depends on the defect strength.
This actually corresponds to two spins located symmetrically about
the defect line, and is related to the one-point function of the
bosonic operator $\cos{\varphi}$.
In general, $\langle \sigma_1 \sigma_2 \rangle$ is smaller than
$\langle \sigma_1 \sigma_1 \rangle$ for the same defect strength
and the same (horizontal) distance, as expected.
Nevertheless, they asymptotically converges to the same power-law
function in the large-distance limit for $b \neq 0$.
Not only the exponent but also the constant prefactor is the same.
This somewhat unexpected phenomenon will be further discussed
in the next subsection using the boundary OPE.

\subsection{Boundary OPE of spin operators}
\label{subsec:boundaryOPE}

As discussed in Ref.~\cite{CardyLewellen},
once we obtain the boundary correlation function
from the bulk OPE,
we can calculate the boundary OPE from the boundary correlation function.

The boundary OPE is the expansion about the boundary limit, where two
operators are close to the boundary (real axis).
Then the cross-ratio $x$, defined in~(\ref{eq:crossratio}),
approaches $1$.
Thus we should expand the correlation function in $\tilde{x} = 1 -x$.
If we define $\tilde{u} = e^{2 \pi i \tilde{\tau}}$ by
\begin{equation}
  \tilde{x} = \left( \frac{\vartheta_2(\tilde{u})}{\vartheta_3(\tilde{u})} \right)^4,
\end{equation}
$\tilde{u}$ is related to $u=e^{2 \pi i \tau}$ by the modular transformation
\begin{equation}
  \tilde{\tau} = - \frac{1}{\tau} .
\end{equation}

The correlation functions~(\ref{eq:boundary11}) and~(\ref{eq:boundary12}) 
can be expressed in terms of $\tilde{u}$
by the modular transformation of the elliptic theta function as
\begin{eqnarray}
  \langle \sigma_1 (z_1,\bar{z}_1) \sigma_1 (z_2,\bar{z}_2) \rangle &=&
       \left( \frac{1}{4 y_1 y_2 x} \right) ^{1/8}
       \frac{\sqrt{2} \tilde{u}^{(\frac{\varphi_0}{\pi})^2}}{
                  \vartheta_3(\tilde{u})}
          \vartheta_3(\tilde{u}^{-2 \varphi_0 / \pi}, \tilde{u}^2), 
\label{eq:boundary11mod}
\\
  \langle \sigma_1 (z_1,\bar{z}_1) \sigma_2 (z_2,\bar{z}_2) \rangle &=&
       \left( \frac{1}{4 y_1 y_2 x} \right) ^{1/8} 
          \frac{\sqrt{2} \tilde{u}^{(\frac{\varphi_0}{\pi})^2}}{
                 \vartheta_3(\tilde{u})}
          \vartheta_4(\tilde{u}^{-2 \varphi_0 / \pi}, \tilde{u}^2) .
\label{eq:boundary12mod}
\end{eqnarray}

From these expressions, we can read off the boundary OPE of $\sigma_1$
and $\sigma_2$.
In particular, for both $\langle \sigma_1 \sigma_1 \rangle$
and $\langle \sigma_1 \sigma_2 \rangle$,
the dimension of the intermediate boundary primary
operators are given by
\begin{equation}
  \Delta_b = 2n^2 + \frac{4 n \varphi_0}{\pi} + 2 \frac{{\varphi_0}^2}{\pi^2}
\end{equation}
where $n$ is an arbitrary integer.
They are exactly the boundary scaling dimensions of the primaries
in the ``moving sector''~(\ref{eq:movingsector}).
Thus the ``moving sector'' is identified as the set of boundary operators
generated by the boundary OPE of the spin operators.

We note that, while the intermediate states in the bulk OPE are different
for $\sigma_1 \sigma_1$ and $\sigma_1 \sigma_2$,
the intermediate states in the boundary OPE are common between them.
Only the OPE coefficients are different.
Moreover,
the OPE coefficient for the boundary operator with the smallest
dimension is exactly the same between $\langle \sigma_1 \sigma_1 \rangle$
and $\langle \sigma_1 \sigma_2 \rangle$.
This means that the asymptotic behavior of these two correlation functions
at large distance is exactly the same.
(For negative $b$, they differ only by their sign.)
The only exception is the free boundary case ($b=0$).
In this case, there are two boundary operators for each boundary dimension
and the two coefficients cancel exactly to make
$\langle \sigma_1 \sigma_2 \rangle$ identically zero.

The boundary OPE coefficients for $\sigma_1 \pm \sigma_2$ are
given by the sum of (or the difference between) those for $\sigma_1$
and $\sigma_2$.
The smallest boundary dimension is given by $2 (\frac{\varphi_0}{\pi})^2$
and $2 ( 1 - \frac{\varphi_0}{\pi})^2$ respectively for
$\sigma_1 + \sigma_2$ and $\sigma_1 - \sigma_2$.
As discussed in Section~\ref{subsec:spectrum},
they are the only relevant boundary operators for this boundary condition.

\subsection{Correlation functions of the bosonic operators}

Here we consider correlation functions of the bosonic operators
$\cos{k \varphi}$.
For $k=1$, $\sqrt{2} \cos{ \varphi}$ is identified with
the composite spin operator $\sigma_1 \sigma_2$ (at the same point).
Thus the boundary two-point function
of  $\cos{ \varphi}$ corresponds to a special case of the four-spin
correlation function
in the presence of the defect line:
correlation function of two pairs, each of which consists of two
spins located symmetrically about the defect line.

The boundary one-point function can be determined by the
method of Ref.~\cite{CardyLewellen} as
\begin{equation}
  \langle \cos{ k \varphi} \rangle =
           \frac{\cos{k \varphi_0}}{(2y)^{1/4}} .
\label{eq:cos1p}
\end{equation}
In order to calculate the boundary two-point function
$\langle \cos{k \varphi} \cos{k \varphi} \rangle$,
we use a kind of mirror image method.
Since it is symmetric under $\varphi \rightarrow - \varphi$,
it is sufficient to consider only the fixed boundary condition
$\varphi = \varphi_0$ at the boundary, apart from an ambiguity in the
overall constant.
The overall constant can not be determined by the mirror image method
and will be fixed later by requiring the correct bulk limit.

Given the boundary condition $\varphi = \varphi_0$ at the boundary
(real axis), the non-chiral boson field $\varphi(z,\bar{z})$ can be
represented by a single chiral boson $\phi(z)$ as
\begin{equation}
  \varphi(z,\bar{z}) = \varphi_0 + \phi(z) - \phi(z'),
\end{equation}
where $z'$ is the complex conjugate of $z$.
Thus the boundary two-point function is mapped to the chiral four-point
function as
\begin{eqnarray}
\lefteqn{  \langle \cos{k\varphi(z_1,\bar{z}_1)}
      \cos{k\varphi(z_2,\bar{z}_2)} \rangle
= }
\nonumber \\
&& \frac{1}{4} \sum_{\pm \pm}
  \langle  e^{ \pm i k [\varphi_0 + \phi(z_1) - \phi(z'_1)]}
           e^{ \pm i k [\varphi_0 + \phi(z_2) - \phi(z'_2)]}
  \rangle ,
\end{eqnarray}
where $\pm$'s are summed independently.

Using the standard result on the correlation functions of the vertex operators
\begin{equation}
  \langle \prod_j e^{i \alpha_j \phi(z_j) } \rangle =
           \prod_{j < k } (z_j - z_k)^{\alpha_j \alpha_k /4} 
\;\;\;\;\;\;
(\sum_j \alpha_j = 0) ,
\label{eq:vertexops}
\end{equation}
this reduces to
\begin{equation}
  \langle \cos{k\varphi(z_1,\bar{z}_1)}
      \cos{k\varphi(z_2,\bar{z}_2)} \rangle
= \frac{C}{(4y_1 y_2)^{k^2/4}}
       [ x^{-k^2/4} + x^{k^2/4} \cos{2 k \varphi_0} ]
\label{eq:coscos}
\end{equation}
where $C$ is a constant and $y_1$, $y_2$ and $x$ are defined
similarly with eq.~(\ref{eq:boundary11}).
Requiring the correct bulk limit
\begin{equation}
  \langle \cos{k\varphi(z_1,\bar{z}_1)}
      \cos{k\varphi(z_2,\bar{z}_2)} \rangle
  \rightarrow \frac{1}{2 |z_1 - z_2 |^{k^2/4}} , 
\end{equation}
$C$ is fixed to be $1/2$.

Let us compare our result~(\ref{eq:coscos})
for $k=1$, which is the special case
of four-spin correlation function, with known results.
For periodic ($\varphi_0 = \pi/4$)
and antiperiodic ($\varphi_0 = 3 \pi/4$) boundary conditions,
our result agrees with the bulk four-point spin correlation
of the Ising model for the special location of the spins.
For the free ($\varphi_0 = \pi/2$) boundary condition,
our result agrees with the square of
the boundary two-spin correlation~\cite{Cardy:BCFT}
for the free boundary condition.
Again we found consistency of our approach with the known results.

\subsection{Two-spin correlation near the end of the defect line}

\begin{figure}[htbp]
  \begin{center}
    \leavevmode
    \epsfxsize=\textwidth
    \epsfbox{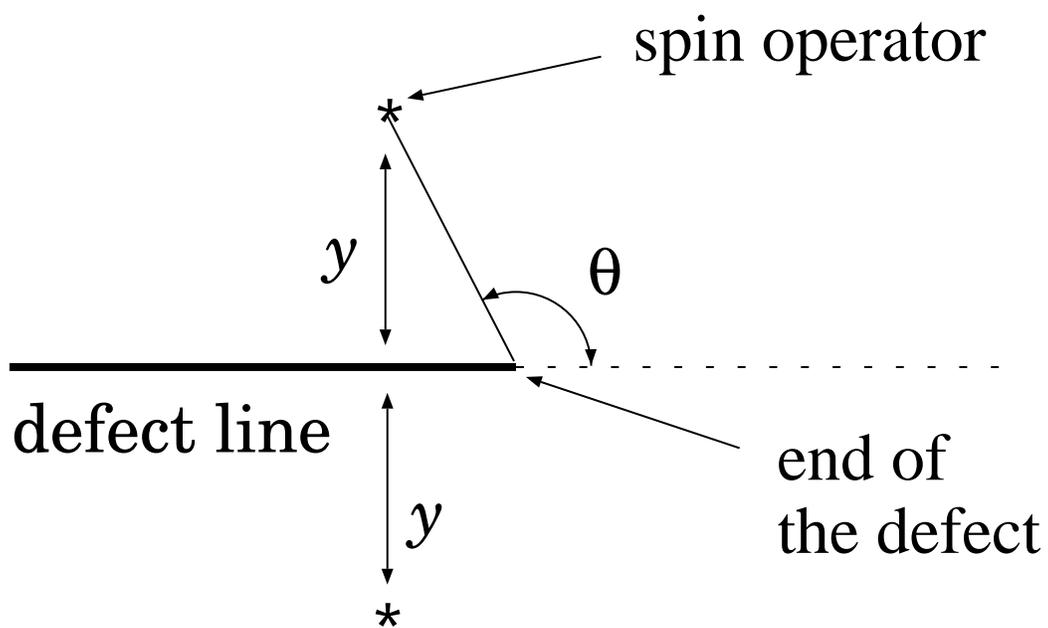}
    \caption{The correlation function of two spins
located symmetrically about the defect line.
It is characterized by the distance from the line $y$ and
the angle $\theta$.}
    \label{fig:enddefect}
  \end{center}
\end{figure}

So far we have discussed only the defect line without ends
(infinitely long or periodic).
Let us consider the neighborhood of the end of the defect line.
We discuss only the correlation function of two spins located
symmetrically about the defect line, as shown in Fig.~\ref{fig:enddefect}.
This can be reduced to the one-point function
$\langle \sqrt{2} \cos{\varphi} \rangle = \langle \sigma_1 \sigma_2 \rangle$
after the folding, with a changing boundary condition.

We regard the end point as the origin, and make the conformal
mapping $z = e^{w}$ to the strip with different boundary conditions
at two sides.
Then the one-point function is given by the vacuum expectation value
with two specified boundary conditions.
The Hilbert space of the strip is classified into several sectors
by the zero-mode. 
We only need to consider the lowest energy zero mode to
calculate the vacuum expectation value.
Thus we fix $\varphi = \varphi_0$ at the boundary corresponding
to the defect and $\varphi = \pi/4$ at the other boundary.

We employ the mirror image method used in the last subsection.
In order to ensure the boundary conditions on the two boundaries,
we first need two mirror images as
\begin{equation}
  \varphi(w, \bar{w}) \sim
       \frac{\pi}{4} + (\varphi_0 - \frac{\pi}{4})\frac{v}{\pi}
               + \phi(w) -  \phi(w') - \phi(w' + 2 \pi i),
\end{equation}
where $\phi$ is a chiral boson,
$w'$ is the complex conjugate of $w$ and $v = {\rm Im} w$.
However, the mirror image about one boundary conflicts with the
boundary condition on the other boundary.
Thus we need the mirror image of the images.
In this way, we need infinitely many mirror images as
\begin{equation}
  \varphi(w, \bar{w}) =
       \frac{\pi}{4} + (\varphi_0 - \frac{\pi}{4})\frac{v}{\pi}
                   - \sum_n  \phi(w' + 2\pi n i)
                   + \sum_n \phi(w + 2 \pi n i).
\end{equation}
Using eq.~(\ref{eq:vertexops}), the one-point function is given by
\begin{eqnarray}
 \langle \cos{\varphi} \rangle &= &
\mbox{const.}
\cos{[ \frac{\pi}{4} + (\varphi_0 - \frac{\pi}{4})\frac{v}{\pi} ]}
\left( \prod_{n = -\infty}^{\infty} \frac{2 \pi n}{2v + 2 \pi n}
\right) ^{1/4}
\nonumber \\
&=&
\mbox{const.} 
\frac{\cos{[ \frac{\pi}{4} + (\varphi_0 - \frac{\pi}{4})\frac{v}{\pi} ]}}{
      (\sin{v})^{1/4}} .
\end{eqnarray}
Going back to the half plane, we obtain the one-point function, or
the special two-spin correlation before folding as in Fig.~\ref{fig:enddefect},
near the end of the defect line.
It is given by
\begin{equation}
\langle \sigma_1 (z,\bar{z}) \sigma_2 (z,\bar{z}) \rangle =
\sqrt{2}
\frac{\cos{[ \frac{\pi}{4} + (\varphi_0 - \frac{\pi}{4})\frac{\theta}{\pi} ]}}{
      (2y)^{1/4}}, 
\end{equation}
where $z=0$ is the end of the defect line,
$y = {\rm Im} z$, $\theta = \arg{z}$, and the overall normalization
is determined from the $\theta \rightarrow 0 , \pi$ limit.
In the $\theta \rightarrow 0$ limit, the spins are located far apart
from the defect line and the correlation function should approach
the bulk value $(2y)^{-1/4}$.
In the $\theta \rightarrow \pi$ limit, the spins are located
symmetrically about the defect line and
are far apart from the end of the defect.
Thus the correlation function should converge to that for the
infinitely long defect~(\ref{eq:cos1p}).

\subsection{Free energy of the finite length defect line}

Here we discuss the free energy due to the presence of the defect line
of the continuous Dirichlet universality class~(\ref{eq:bstateAT}).
A defect line with a finite length corresponds to a finite interval
with a different boundary condition, after the folding.
The partition function in the presence of the defect line
is expressed by the two-point function of a boundary condition
changing operator~\cite{Cardy:fusion}.
(For further discussion, see Appendix.)

The dimension of the boundary condition changing operator
is determined by the ground-state energy
of a strip with boundary conditions corresponding to the defect line
and no defect ($b=1$) on the two sides.
From the partition function~(\ref{eq:totalZ}),
the dimension is found as
\begin{equation}
  \Delta_{b} = \frac{1}{2}\left[ \frac{1}{\pi}(\varphi_0 - \frac{\pi}{4})
                           \right] ^2 .
\end{equation}
The universal term in the free energy of a defect of length $l$
on an infinite plane is thus given by
\begin{equation}
  F_{\rm univ}(l) = - k T \log{Z} =  k T \Delta_{b} \log{l} .
\end{equation}
The universal free energy
of the defect line along the circumference $L$ of
a  cylinder can be obtained by a conformal mapping $z = \tan{\pi w/L}$ as
\begin{equation}
 F_{\rm univ}(l,L) =
   k T \Delta_{b} \log{\left[ \frac{L}{\pi}  \sin{\frac{\pi l}{L}} \right]} .
\end{equation}

However, the above is not the only term in the free energy.
There are also non-universal terms.
Firstly, there is a non-universal shift of the ground state energy
depending on the boundary condition, as seen in the second term
in eqs.~(\ref{eq:oddvac}) and~(\ref{eq:evenvac}).
This is due to the phase shift of the fermion momenta~(\ref{eq:qcond}),
which depends on the fermion energy.
This effect, which is known as Fumi's theorem~\cite{Mahan:text}
in the general case,
depends on the dispersion relation and the phase shift in the
entire Brillouin zone, and thus is non-universal.
In general, we expect a shift of the ground-state energy of order $1/a$
and the free energy proportional to the defect length as $A l$,
where $A$ is a constant of order $1/a$.
An exception is the antiferromagnetic defect ($b=-1$).
In this case, the quantization of the wavevectors is exactly the same
as in the no defect case and only the parity assignment is reversed.
Therefore the non-universal term proportional to the length is absent
($A=0$).
This is also required from the fact that the antiferromagnetic defect
is exactly the two-point correlation of disorder operators~\cite{KadanoffCeva},
which decays purely algebraically.

Secondly, the free energy of the defect line should vanish
when there is no defect ($l=0$).
Hence a constant term of order $\log{a}$, where $a$ is the cutoff length
(lattice constant) is required.
The combined result would be approximately
$k T \Delta_b \log{\frac{l}{a}}$, which is only valid for $l >> a$.
The constant term of order $\log{a}$ is non-universal and can be regarded
as an effect of the ends of the defect line (or, of the insertion of the
boundary condition changing operator.)
Moreover, the dependence of the free energy on $l$ when $l \sim a$
should be non-universal.
If the defect length approaches to the circumference
of the cylinder ($l \sim L$),
the dependence on $l$ becomes again non-universal for $(L - l) \sim a$.
When $l = L$, however, the defect line completely wraps around the
cylinder; there is no longer a non-universal constant due to the
ends of the defect line.
Nevertheless, a universal constant of order $1$ still remains in general.
After the folding, we have a new single boundary condition.
The universal remaining constant is nothing but the contribution
from the ``ground-state degeneracy''~\cite{Affleck:gtheorem}:
$-kT \log{g}$.
In the present problem, $g = 1$ for any continuous Dirichlet boundary
state, as will be discussed in Section~\ref{sec:RGflow}.
Thus the universal constant for $l = L$ is zero.

As discussed above,
the non-universal part in the free energy of the defect line
is rather large except for $b=-1$; it may be difficult to measure the
universal part in (numerical) experiments.
(The dimensions of boundary condition changing operators are measurable
in quantum impurity problems~\cite{Affleck:bcchange}.)

\subsection{Duality and disorder correlation}
\label{subsec:dual}

Our model~(\ref{eq:defectham}) has a defect line where
the strength of vertical
(i.e. orthogonal to the defect line) links are altered.
Under the duality transformation~\cite{KramersWannier,FradkinSusskind},
our model~(\ref{eq:defectham})
transforms to the Ising model 
with the defect line where the strength of the horizontal coupling is altered.
This ``horizontal'' model was studied by several
authors~\cite{Bariev:Ising,McCoyPerk,Kadanoff:defect,Brown:defect}.
The duality transformation exchanges the order and disorder operator.
It is illuminating to discuss the duality transformation
in the quantum model under the $\tau$-continuum limit.
Our ``vertical'' model corresponds to the
quantum Hamiltonian~(\ref{eq:qham}).
The duality transformation~\cite{FradkinSusskind}
is introduced by defining the new operators
$\hat{\mu}$ as in
\begin{equation}
\begin{array}{rcl}
  \hat{\mu}^z(n) & \equiv  & \prod_{ m \leq n} \hat{\sigma}^x(m) ,
\\
  \hat{\mu}^x(n) & \equiv & \hat{\sigma}^z(n) \hat{\sigma}^z(n+1),
\end{array}
\label{eq:dualtr}
\end{equation}
where $\hat{\sigma}$'s are defined as in~(\ref{eq:qham}). 
$\hat{\mu}^z$ is the disorder operator.

We apply this dual transformation for $- \infty < n < \infty$.
Using the new operators $\hat{\mu}$,
the Hamiltonian~(\ref{eq:qham}) is expressed as
\begin{equation}
\label{eq:dualqham}
  H = - \sum_{n \neq 0 } \hat{\mu}^x(n)
      - \sum_{n = -\infty}^{\infty} \hat{\mu}^z(n-1) \hat{\mu}^z(n)
      - b \hat{\mu}^x(0) .
\end{equation}
Now the Hamiltonian is again the transverse Ising model, but
the only the strength of the transverse field is altered at $n=0$.
This corresponds to the variation of the horizontal link strength.

Let us consider the model by a perturbation from the model without
the defect ($b=1$) as in Ref.~\cite{Brown:defect}.
In the continuum limit,
the variation of either vertical or horizontal coupling correspond
to the perturbation by the same energy density operator.
Thus, if the continuum perturbation theory is valid in principle,
the ``horizontal'' model is equivalent to the ``vertical''
model~(\ref{eq:defectham}) under some change of the parameter
$\varphi_0$.
Assuming this, the Dirichlet boundary state is transformed
to the Dirichlet boundary state with a different parameter
$\varphi_0$ under the duality transformation.
The transformation rule for the parameter $\varphi_0$
can be determined from Brown's relation~\cite{Brown:defect}
between the surface exponents of the order/disorder fields:
\begin{equation}
  \sqrt{ 2 \Delta_b (\sigma) } + \sqrt{ 2 \Delta_b(\mu)} =1 .
\end{equation}
This relation together with our result~(\ref{eq:boundaryexp})
gives the transformation of $\varphi_0$ under the
duality transformation:
\begin{equation}
  \varphi_0 \rightarrow
  \varphi_0^{d} = 
     \left\{ \begin{array}{lr}
       \frac{\pi}{2} - \varphi_0 & ( 0 < \varphi_0 < \frac{\pi}{2} ) \\
       \varphi_0 - \frac{\pi}{2} & ( \frac{\pi}{2} < \varphi_0 < \pi) 
     \end{array} \right. .
\label{eq:dualphidef}
\end{equation}
Thus the correlation functions of two disorder operators in the 
Dirichlet boundary state are given by
\begin{equation}
  \langle \mu_j \mu_k \rangle_{D (\varphi_0)} =
  \langle \sigma_j \sigma_k \rangle_{D (\varphi_0^{d})}
\end{equation}

\section{Properties of the continuous von Neumann universality class}
\label{sec:Ncorr}

Here we discuss the correlation functions near the
defect line that is identified with the
continuous  von Neumann boundary state~(\ref{eq:orbNeuBS}).
We also identify the corresponding defect line in the one-dimensional
quantum model.

\subsection{Two-spin correlation functions}

We can calculate the two-spin correlation functions
$\langle \sigma_i \sigma_j \rangle$ in a similar manner
to what we did in Section~\ref{subsec:spincorr}.
The only difference is the boundary state, and consequently
the coefficients $A^{\Delta}_B$ defined in~(\ref{eq:boundary1pt}).
The boundary state~(\ref{eq:orbNeuBS}) determines the coefficients as
\begin{equation}
  A^{(n^2/2, A)}_B = \left\{  \begin{array}{ll}
                              1 & (n=0) \\
                      \sqrt{2} \cos{(2 n \tilde{\varphi}_0)} & (n \geq 1) 
                            \end{array} \right.  
\end{equation}

Firstly, it is easy to see that $\langle \sigma_1 \sigma_2 \rangle$
vanishes for any value of $\tilde{\varphi}_0$, since the bulk OPE
$\sigma_1$ and $\sigma_2$ only generates the operators
as in~(\ref{eq:dimOPE12}).
 
The bulk OPE coefficients of $\sigma_j \sigma_j$ ($j=1,2$) are given by
\begin{eqnarray}
C_{\sigma_1 \sigma_1 (n^2/2, A)} &=&  
\sqrt{2} \left( \frac{1}{16} \right)^{n^2/2} ,\\
C_{\sigma_2 \sigma_2 (n^2/2, A)} &=&  
(-1)^n \sqrt{2} \left( \frac{1}{16} \right)^{n^2/2} .
\end{eqnarray}
Thus the correlation functions are given by
\begin{eqnarray}
\lefteqn{  \langle \sigma_1 (z_1,\bar{z}_1) \sigma_1 (z_2,\bar{z}_2) \rangle }
\nonumber \\
&=& \left( \frac{1}{4 y_1 y_2 x} \right)^{1/8} \frac{1}{\vartheta_3(u)}
   \sum_{n=0}^{\infty} A^{(n^2/2,A)}_B C_{\sigma_1 \sigma_1 (n^2/2,A)}
                           f(\frac{n^2}{2},\frac{1}{16},1,x)
\nonumber \\
&=& 
\left( \frac{1}{4 y_1 y_2 x} \right)^{1/8} \frac{1}{\vartheta_3(u)}
 ( 1 + \sum_{n=1}^{\infty} 2 \cos{(2 n \tilde{\varphi}_0)} u^{\frac{n^2}{4}} )
\nonumber \\
&=&
\left( \frac{1}{4 y_1 y_2 x} \right)^{1/8} \frac{1}{\vartheta_3(u)}
          \vartheta_3(e^{2 i \tilde{\varphi}_0}, \sqrt{u}), 
\\
\lefteqn{  \langle \sigma_2 (z_1,\bar{z}_1) \sigma_2 (z_2,\bar{z}_2) \rangle }
\nonumber \\
&=& \left( \frac{1}{4 y_1 y_2 x} \right)^{1/8} \frac{1}{\vartheta_3(u)}
   \sum_{n=0}^{\infty} A^{(n^2/2,A)}_B C_{\sigma_2 \sigma_2 (n^2/2,A)}
                           f(\frac{n^2}{2},\frac{1}{16},1,x)
\nonumber \\
&=& 
\left( \frac{1}{4 y_1 y_2 x} \right)^{1/8} \frac{1}{\vartheta_3(u)}
 ( 1 + \sum_{n=1}^{\infty} 2
     (-1)^n \cos{(2 n \tilde{\varphi}_0)} u^{\frac{n^2}{4}} )
\nonumber \\
&=&
\left( \frac{1}{4 y_1 y_2 x} \right)^{1/8} \frac{1}{\vartheta_3(u)}
          \vartheta_4(e^{2 i \tilde{\varphi}_0}, \sqrt{u}) .
\end{eqnarray}
They are no longer symmetric under the exchange of $\sigma_1$ and
$\sigma_2$, or equivalently the reflection about the defect line
before the folding.
We notice that these correlation functions are actually identical
to those for the continuous Dirichlet
universality class~(\ref{eq:boundary11},\ref{eq:boundary12})
for an appropriate value of the parameter.
We summarize this observation below
\begin{equation}
\begin{array}{rcl}  
  \langle \sigma_1 \sigma_1 \rangle_{N ( \tilde{\varphi}_0 )} &=&
  \langle \sigma_1 \sigma_1 \rangle_{D ( \varphi_0 = \tilde{\varphi}_0 )}
\\
\langle \sigma_2 \sigma_2 \rangle_{N ( \tilde{\varphi}_0 )} &=&
\langle \sigma_1 \sigma_1 \rangle_{D ( \varphi_0 = \pi/2 - \tilde{\varphi}_0 )}
\\
 \langle \sigma_1 \sigma_2 \rangle_{N ( \tilde{\varphi}_0 )} &=&
0 ,
\end{array}
\label{eq:Neucorr}
\end{equation}
where $\langle \rangle_{N ( \tilde{\varphi}_0 )}$ and
$\langle \rangle_{D ( \varphi_0 )}$ are the correlation function
in the von Neumann and Dirichlet boundary states, respectively.
It is also possible to calculate other quantities discussed in
Section~\ref{sec:Dcorr}
for the continuous von Neumann universality class.

\subsection{Identification of the defect line}

Naively the result $\langle \sigma_1 \sigma_2 \rangle =0$
implies that the two layers are completely decoupled.
If they are completely decoupled, the boundary state must be
$| ff \rangle $ or equivalently the Dirichlet one
with $\varphi_0 = \pi/2$. 
However, for example, at $\tilde{\varphi}_0 = \pi/4$ the correlation functions
$\langle \sigma_1 \sigma_1 \rangle$ and $\langle \sigma_2 \sigma_2 \rangle$
behaves exactly as if there is no defect, while
$\langle \sigma_1 \sigma_2 \rangle$ still vanishes.
These behaviours are definitely different from those in $| ff \rangle$
boundary states.
Hence there should be some sort of coupling between two layers.

We have not yet found an appropriate defect line in the classical model
which corresponds to the continuous von Neumann boundary state.
However, we have succeeded in finding an appropriate point defect
in the corresponding one-dimensional quantum model.
It is given by the Hamiltonian:
\begin{equation}
\label{eq:qham2}
  H = - \sum_{n \neq 0} \hat{\sigma}^x(n)
      - \sum_{n \neq 0} \hat{\sigma}^z(n-1) \hat{\sigma}^z(n)
      - b \hat{\sigma}^z(-1) \hat{\sigma}^x(0), 
\end{equation}
where the operators are defined similarly as in eq.~(\ref{eq:qham}).

We show that this model is related to the previous model~(\ref{eq:qham})
by a duality transformation~(\ref{eq:dualtr})
{\em on one side of the defect only}.
The duality transformation on the half-line $n \geq 0$ is defined by
\begin{eqnarray}
  \hat{\mu}^z(n) & \equiv  & \prod_{0 \leq m \leq n} \hat{\sigma}^x(m) ,
\\
  \hat{\mu}^x(n) & \equiv & \hat{\sigma}^z(n) \hat{\sigma}^z(n+1).
\end{eqnarray}
The new operators $\hat{\mu}$'s are defined only for $n \leq 0$ and
commute with $\hat{\sigma}$'s at $n < 0$.
They obey the same commutation relation as the original $\hat{\sigma}$'s do.
The Hamiltonian~(\ref{eq:qham2}) is rewritten as
\begin{eqnarray}
H &=& - \sum_{n <0 } \hat{\sigma}^x(n) - \sum_{n \geq 0} \hat{\mu}^x (n)
\nonumber \\
&& 
- \sum_{n < 0} \hat{\sigma}^z(n-1) \hat{\sigma}^z(n)
- \sum_{n > 0} \hat{\mu}^z(n-1) \hat{\mu}^z(n)
\nonumber \\
&&
- b \hat{\sigma}^z(-1) \hat{\mu}^z(0) .
\end{eqnarray}
This is exactly the Hamiltonian~(\ref{eq:qham}),
under the redefinition $\hat{\mu} \rightarrow \hat{\sigma}$.

In this way, the correlation functions in our model $\langle \rangle_M$
are identified
with the correlation functions in the Dirichlet boundary condition
as follows:
\begin{eqnarray}
  \langle \sigma_1 \sigma_1 \rangle_{M} &=&
  \langle \sigma_1 \sigma_1 \rangle_{D ( \varphi_0 )} ,\\
  \langle \sigma_2 \sigma_2 \rangle_{M} &=&
  \langle \mu_2 \mu_2 \rangle_{D ( \varphi_0 )}  \nonumber \\
&=& 
  \langle \sigma_1 \sigma_1 \rangle_{D ( \varphi_0^{d} )} ,\\
  \langle \sigma_1 \sigma_2 \rangle_{M} &=&
  \langle \sigma_1 \mu_2 \rangle_{D ( \varphi_0 )} =0  ,\\
\end{eqnarray}
where $\mu$ is the disorder operator~\cite{KadanoffCeva},
$\varphi_0^{d}$ is defined in~(\ref{eq:dualphidef}) and
$\varphi_0$ is determined by the parameter $b$ as in~(\ref{eq:phi0def}).

These results exactly agree with the correlation functions~(\ref{eq:Neucorr})
in the von Neumann boundary condition
if $\varphi_0 = \tilde{\varphi}_0$.
Therefore our model~(\ref{eq:qham2}) presumably corresponds to
the continuous von Neumann boundary state~(\ref{eq:orbNeuBS})
in the continuum limit.

\section{Renormalization group flow among the boundary states}
\label{sec:RGflow}

\subsection{The relative stability of the boundary states}

Now we discuss the relative stability and the renormalization group
flow among the boundary states.
In this section, the orbifold radius $r$ is again fixed to $1$
as we discuss the Ising defect problem.
The universal ``ground-state degeneracy''
$g$~\cite{Affleck:gtheorem}, which is given by the
inner product of the boundary state and the
ground state, turns out to be $1$ in the continuous Dirichlet boundary
state~(\ref{eq:bstateboson})
for any value of $\varphi_0$.
Similarly, $g=\sqrt{2}$ in the continuous von Neumann
boundary state~(\ref{eq:orbNeuBS})
independent of $\tilde{\varphi}_0$.
The number $g$ plays the role similar to the central charge $c$ in the
bulk conformal field theory.
According to the ``$g$-theorem''~\cite{Affleck:gtheorem},
$g$ decreases along a renormalization group trajectory connecting
two conformally invariant boundary conditions.

In~(\ref{eq:bstateboson}) and~(\ref{eq:orbNeuBS}),
the boundary state depends continuously on the
defect strength $b$, namely there is a fixed line of
conformally invariant boundary conditions.
Thus it is natural that $g$ is constant along two continuous line
of boundary universality class, as central charge $c$ is constant
along a fixed line of conformally invariant bulk theories.
Namely, no renormalization group flow can occur along these lines.
The values of $g$ in the 8 discrete boundary states are simply the
product of $g$ for the two Ising layers.
It is known~\cite{Cardy:fusion} that
$g=1$ for the free Ising boundary condition and $g=1/\sqrt{2}$
for the fixed Ising boundary condition.
Thus $g$ is $1/\sqrt{2}$ for the endpoint Neumann boundary states
($| \uparrow f \rangle $ etc.) and $1/2$ for the endpoint Dirichlet
boundary states ($| \uparrow \uparrow \rangle$).

By the ``$g$-theorem'',
the continuous Neumann boundary state is the most unstable among
these boundary states.
The next most unstable one is the continous Dirichlet, then
the endpoint Neumann, and the most stable ones are the endpoint Dirichlet.
If there is no restriction on the defect line, it is thus
generically attracted to one of the endpoint Dirichlet boundary states.
In order to realize the other universality classes, we must protect
the defect line from the relevant operators, for example by
an appropriate symmetry.
From the expression of the boundary states as the tensor products
of the Ising boundary states,
it is clear that the discrete endpoint boundary states break the
global $Z_2$ symmetry: $\sigma \rightarrow - \sigma$.
Thus if we do not break the global $Z_2$ symmetry,
the defect line is renormalized to one of the continuous states.
On the other hand, the endpoint Neumann boundary states are expected
to have a different $Z_2$ symmetry: Ising spin flip only on the one side
of the defect line.
For example, $| \uparrow f \rangle$ is invariant under the
Ising spin flip in the second layer.
Therefore the defect line will be renormalized into one of the endpoint
Neumann ones if we break the global $Z_2$ symmetry but preserve
the other $Z_2$ symmetry.

Analysis of the operator contents is useful to further investigation of
this problem.
We can obtain the spectrum of the boundary operators in
the continuous Neumann boundary states
in a similar manner as in Section~\ref{subsec:spectrum}.
The partition function on the strip with the same Neumann boundary
on both sides is given by
\begin{equation}
  Z = \frac{1}{\eta(q)} \sum_{n = - \infty}^{\infty} q^{n^2/2}
     + \frac{1}{\eta(q)}
  \sum_{n = - \infty}^{\infty} q^{(n + 2\tilde{\varphi}_0/\pi)^2/2} ,
\end{equation}
where $q$ is defined as in~(\ref{eq:qandw}).

From this expression, we can read off the whole spectrum of boundary
operators.
In particular, the dimensions of the relevant operators
(other than the identity) are given by
$1/2$ (twice), $2 ( \frac{\tilde{\varphi}_0}{\pi} )^2$,
$ \frac{1}{2} ( 1 - \frac{ 2 \tilde{\varphi}_0}{\pi})^2$.
From the two-spin correlation functions~(\ref{eq:Neucorr}),
the latter two are identified with the spin operators $\sigma_1$ and
$\sigma_2$.
The remaining two with dimension $1/2$ are identified with
$\sigma_1 \sigma_2$.
The multiplicity $2$ is understood as follows.
Since the dimension $1/2$ is independent of the parameter
$\tilde{\varphi}_0$, we may consider the simplest case $b=1$
($\tilde{\varphi}_0 = \pi/4$).
At this point, the problem is reduced to the Ising model without
any defect after the duality transformation.
Now we can unfold the system and reduce
the boundary OPE between $\sigma_1$ and $\sigma_2$ to the
bulk OPE between $\sigma$ and $\mu$, which is given~\cite{KadanoffCeva,BPZ} as
\begin{equation}
  \sigma(w,\bar{w}) \mu(0,0) \sim
              w^{3/8} \bar{w}^{-1/8} \psi +
              w^{-1/8} \bar{w}^{3/8} \bar{\psi} .
\end{equation}
Thus there are actually two operators $\psi$ and $\bar{\psi}$
generated by the OPE of $\sigma$ and $\mu$.
The coefficients depend on the relative location of $\sigma$ and $\mu$.
Translating to the folded picture, two boundary operators are
generated by the OPE of $\sigma_1$ and $\sigma_2$
and the relative coefficients depend on the location of $\sigma_1$
and $\sigma_2$.
While the above argument is only valid for the special point $b=1$,
it is natural to expect that two boundary operators are generated
by the OPE for any value of $b$ (or $\tilde{\varphi}_0$).

Now let us turn to the stability of the Neumann boundary state.
Since the spin operators $\sigma_1$ and $\sigma_2$ are relevant,
the perturbation by these operators will drive the system away from
the von Neumann boundary state.
The resulting state will be one of the discrete boundary states.
For example, if we perturb by $\sigma_1$, the boundary will be
presumably renormalized to $| \uparrow f \rangle$
(or $ | \downarrow f \rangle$)
boundary state.
The remaining relevant operators are generated by $\sigma_1 \sigma_2$.
Under this perturbation,
the boundary will be presumably renormalized to the Dirichlet
boundary state with some value of $\tilde{\varphi}_0$.
These conclusions are
consistent with our discussion based on the ``$g$-theorem''.

To achieve the continuous Neuman boundary state in the infrared limit,
we must protect the boundary from these relevant operators.
The model~(\ref{eq:qham2}) has the following $Z_2 \times Z_2$ symmetry: 
\begin{equation}
\begin{array}{ccc}
  \hat{\sigma}^z(n) \rightarrow - \hat{\sigma}^z (n) & ( n < 0)
& \;\;\; \mbox{with} \;\; \hat{\sigma}^x(0) \rightarrow - \hat{\sigma}^x(0)
\\
  \hat{\sigma}^z(n) \rightarrow - \hat{\sigma}^z (n) & ( n \geq 0 )
\end{array}
\label{eq:Z2Z2inNeu}
\end{equation}
In the continuum limit, it may be simply interpreted as
\begin{equation}
 \sigma_1 \rightarrow - \sigma_1 \;\; \mbox{and} \;\;
 \sigma_2 \rightarrow - \sigma_2 .
\end{equation}
Any of the relevant operators $\sigma_1$, $\sigma_2$ and
$\sigma_1 \sigma_2$
breaks this $Z_2 \times Z_2$ symmetry.
This means that
the $Z_2 \times Z_2$ symmetry protects the von Neumann boundary
state.
However, the Dirichlet boundary state with $\varphi_0 = \pi/2$
(i.e. $| ff \rangle$) also has the $Z_2 \times Z_2$ symmetry.
Hence a defect line with the $Z_2 \times Z_2$ symmetry may not
renormalize into the von Neumann boundary state;
an obvious example is our original model~(\ref{eq:defectham}) with $b=0$.

As we have discussed,
the $\sigma_1 \sigma_2$ perturbation will presumably drive
the defect line to
the Dirichlet boundary state, where only the single global $Z_2$ symmetry
$\sigma_{1,2} \rightarrow - \sigma_{1,2}$ exists.
This is natural since $\sigma_1 \sigma_2$ breaks the
$Z_2 \times Z_2$ symmetry to $Z_2$.
On the other hand,
the reflection symmetry $\sigma_1 \leftrightarrow \sigma_2$,
which was absent in the continuous Neumann boundary state,
is recovered in the continuous Dirichlet boundary state.

The boundary operator content in the continuous Dirichlet boundary
state is given in Section~\ref{subsec:spectrum}.
The two boundary operators with the dimensions
$2 (\varphi_0/\pi)^2$ and $2 [(\pi - \varphi_0) / \pi]^2$
are only the relevant boundary operators.
As we have shown in Section~\ref{subsec:boundaryOPE},
they correspond to $\sigma_1 \pm \sigma_2$.
At the ``periodic'' point $\varphi_0 = \pi/4$, 
$\Delta_b = 1/8, 9/8$ which correspond to
$\sigma$ and $d \sigma / d x \sim (\sigma_1 - \sigma_2) / a$.
At the ``free'' point $\varphi_0 = \pi/2$,
$\Delta_b = 1/2 , 1/2$ since $\sigma_1$ and $\sigma_2$ are independent.

Thus, if we preserve the global $Z_2$ symmetry
$\sigma_1 , \sigma_2 \rightarrow - \sigma_1 , - \sigma_2$,
the fixed line is stable.
If we perturb the defect line by the relevant operator
$\sigma_1 \pm \sigma_2$, the boundary state will be presumably
renormalized to one of the magnetized boundary states
($| \uparrow \uparrow \rangle$ etc.)
Such a flow is established at the ``free'' point ($\varphi_0 = \pi/2$),
where the boundary condition of the two independent Ising models
are renormalized.
We expect the renormalization group flow from the entire fixed line
to the discrete magnetized boundary states, consistently with
our arguments based on ``$g$-theorem''.

\section{Generalized defects and universality}

Grimm~\cite{Grimm} considered generalizations of the defect line.
Here we discuss his result from our general consideration of
defect universality classes.

His first (``integrable'') generalization of the defect line is
given by the quantum Hamiltonian
\begin{equation}
  H(\alpha, \phi) = -  \sum_{j=1}^{N-1} \left[
             \sigma_j^x + \sigma_j^z \sigma_{j+1}^z \right]
           -  \left[ 
             \sigma_N^x + \alpha \sigma_N^{\phi} \sigma_1^{\phi}
              \right]
\label{eq:Grimm1}
\end{equation}
where 
\begin{equation}
 \sigma^{\phi} \equiv \cos{\phi} \sigma^z + \sin{\phi} \sigma^y.
\label{eq:Grimmspin}
\end{equation}
(Eq.~(\ref{eq:Grimm1}) corresponds to eq.~(3.1)
in Ref.~\cite{Grimm} but we have changed the notation.)
When $\phi=0$, this reduces to the ``ordinary'' defect described by
eq.~(\ref{eq:qham}).
Grimm obtained the exact spectrum of this model.
Apart from the non-universal shift in the ground-state energy,
the result was identical to the ``ordinary defect'', i.e. continous
Dirichlet boundary state~(\ref{eq:bstateboson}) with some value
of $\varphi_0$.
He noticed that the model~(\ref{eq:Grimm1}) preserves the global
$Z_2$ symmetry (simultaneous rotation about $x$-axis by angle $\pi$
at each site) and thus it belongs to the same universality
class as the ``ordinary'' one~(\ref{eq:qham}).
These are completely consistent with our arguments based on
``$g$-theorem'' and the boundary operator contents.

He further introduced another generalization given by
\begin{equation}
  H(\beta, \psi) = - \sum_{j=1}^{N-1} \left[
             \sigma_j^x + \sigma_j^z \sigma_{j+1}^z \right]
           - \left[ 
             \sigma_N^x + 
             \beta \sigma_N^z (\cos{\psi} \sigma_1^z - \sin{\psi} \sigma_1^x)
             \right],
\label{eq:Grimm2}
\end{equation}
where $0 < \psi < \pi$.
(This corresponds to eq.~(4.1) in Ref.~\cite{Grimm} but again
we changed the notation.)
This model is not exactly solvable and he obtained the spectra
numerically.
He calculated the difference of the universal term in the ground-state
energy between~(\ref{eq:Grimm2}) with $\beta > 0$
and eq.~(\ref{eq:qham}) with $b=0$.
The numerical result was consistent with his conjecture that
the difference is given by $0,1/16$ and $1/2$ (in unit of $ 2 \pi v /N$),
respectively for
$\psi < \pi/2$, $\psi = \pi/2$ and $\psi > \pi/2$ (independent of $\beta$
as long as $\beta > 0$.)
He argued that this means that there are three universality classes
for this model.

Let us reconsider his model~(\ref{eq:Grimm2}) from our viewpoint.
As he noticed, this model no longer keeps the global $Z_2$ symmetry.
From our arguments in the previous subsection, we expect that
the defect line will generically renormalize into one of the endpoint
Dirichlet boundary states.
Actually, we can understand as follows. (Here we restrict to the
case in which $\beta > 0$.)
We expect $\langle \sigma_1^x \rangle > 0$ due to
the transverse field in the Hamiltonian~(\ref{eq:Grimm2}).
Thus $\langle \sigma_N^z \rangle$ is expected to be negative
for all $0 < \psi < \pi$, due to the effective field generated
by $ \sigma_1^x $.
Then $\sigma_1^z$ is affected by the effective field generated
by $\sigma_N^z$.
Depending on the sign of the coupling,
$\langle \sigma_1^z \rangle$ should be negative for $0 < \psi < \pi/2$
and positive for $\pi/2 < \psi < \pi$.
At $\psi = \pi/2$, neither direction is favored for $\sigma_1^z$.

Thus we expect that the corresponding boundary state is
$| \downarrow \downarrow \rangle$, $| \downarrow f \rangle$ and
$| \downarrow \uparrow \rangle$, respectively for
$0 < \psi < \pi /2$, $ \psi = \pi/2$ and $\pi/2 < \psi < \pi$.
From the discussion of the previous subsection,
the endpoint Neumann boundary state $| \downarrow f \rangle$ is
unstable and would be realized only at $\psi = \pi/2$.
Actually, at this point we have the extra $Z_2$ symmetry:
$\sigma^z \rightarrow - \sigma^z$ only on one side of the
defect. This symmetry protects the endpoint Neumann boundary
state, as we argued.
Changing $\psi$ from $\pi/2$ corresponds to the relevant perturbation
and the defect line will  renormalize to the endpoint Dirichlet:
$| \downarrow \downarrow \rangle$ or $| \downarrow \uparrow \rangle$
depending on the direction of the perturbation.
We also note that, although his model~(\ref{eq:Grimm2}) at $\psi=\pi/2$
looks similar to our model~(\ref{eq:qham2})
which corresponds to the continuous von Neumann
boundary state, they are actually different.
The transverse field at site $n=0$ exists in the former, breaking
the $Z_2 \times Z_2$ symmetry~(\ref{eq:Z2Z2inNeu}) to the
single $Z_2$ symmetry.

Grimm calculated the ground-state energy of the model~(\ref{eq:Grimm2})
on a circle.
In our formalism, it corresponds to a strip of width $N/2$, with
the defect boundary state on one side and the ``periodic''
(Dirichlet with $\varphi_0 = \pi/4$) boundary state on the other side.
The partition function of the system is readily derived;
only the untwisted sector of the defect boundary states contributes
to the partition function.
From the partition function, the ground-state energy of the strip
is given by
$1/16, 1/8$ and $9/16$ in the unit of $2 \pi v /N$,
respectively for
$|\downarrow \downarrow \rangle$, $|\downarrow f \rangle$ and
$| \downarrow \uparrow \rangle$.
Following his definition, we subtract the ground-state energy
in the case the defect boundary state is $| ff \rangle$,
to obtain $0, 1/16$ and $1/2$ respectively for those three cases.
Thus our analysis is in complete agreement with Grimm's.
Furthermore, his conjecture based on $c=1/2$ character of the
Virasoro algebra also agrees with our boundary states which
is tensor products of Ising boundary states.

The ``new'' universality class of the defect lines
he obtained was actually equivalent to  boundary conditions
of two independent Ising models. 
Nevertheless, his general discussion on the universality
of the defect lines seems quite correct.

Recently, Karevski and Henkel~\cite{Karevski:S1defect} discussed
another kind of generalization.
They studied a junction of $S=1/2$ and $S=1$ quantum Ising chains.
In the continuum limit, the bulk universality class of the $S=1$ Ising
chain is identical to that of standard $S=1/2$
Ising model~\cite{Hofstetter:S1}.
Thus, when both chains are critical, we expect that
the universality class of the junction is identical to that
of the defect discussed in this paper.
Actually, they numerically studied~\cite{Karevski:S1defect}
the junction with the Ising $Z_2$
symmetry, and found the universal behaviour corresponding to
the continuous Dirichlet boundary condition in our terminology.
This also confirms that the notion of universality applies to the
defect problem.

\section{Summary and Discussion}
\label{sec:summary}

We studied the continuum limit of the
two-dimensional critical Ising model with a defect line.
Folding the system at the defect line, we map the system to
the critical Ashkin-Teller model~\cite{AshkinTeller,Yang:ZofAT}
at the decoupling point with boundaries.
Thus we can classify the universality classes of the defect lines
by the boundary states of a conformal field theory.

Based on the exact partition function,
we identified the boundary state corresponding to arbitrary strength
of the defect, in terms of the Ashkin-Teller Ishibashi states.
Furthermore, we discussed the possible boundary states
in the $Z_2$ orbifold of free boson, which is identified with
the continuum limit of the critical Ashkin-Teller model.
There are two continuous lines of boundary states, and
eight discrete ones 
for general values of the compactification radius $r$.
The special case $r=1$ is applied to our problem of the defect
lines in the Ising model.
It is shown that the obtained orbifold boundary states include all
the known defect lines, and one new one-parameter family universality
class of defect lines corresponding to the orbifold von Neumann boundary
state.
We conjectured that the above boundary states exhausts all the possible
boundary states on $Z_2$ orbifold for generic value of $r$;
if this is true, we have identified the complete set of
universality classes of defect lines in the Ising model.

From the boundary state, we obtained the complete spectrum of boundary
operators.
We calculated the exact boundary two-spin correlation
functions for arbitrary strength of the defect,
employing the method of Cardy and Lewellen~\cite{CardyLewellen} and
Zamolodchikov's solution~\cite{Zam:ATspin}
for $c=1$ conformal block of spin operators.
We also calculated the boundary two-point functions of bosonic operators,
one of which is identified as a special case of the four-spin correlation
function.
In addition, we obtained a special case of the two-spin correlation function
near the end of the defect line, the universal term in the
free energy of the defect line and the correlation function of the
disorder operators.
While we made some assumptions during the calculation, 
our approach, which satisfies
several nontrivial consistency check, appears to be valid.

We also calculated the two-point spin correlation functions
in the new one-parameter family of defect lines, and identified
the corresponding quantum model using the duality
transformation. 
For the 8 discrete universality classes, calculation of the
correlation functions reduces to that in the Ising model with
a boundary, which is already solved~\cite{CardyLewellen}.
Thus we have obtained two-point spin correlation functions
for all defect universality classes found so far (i.e. for all
the universality classes, if our conjecture is correct.)
Finally, we discussed the renormalization-group flow among the
defect universality classes based on the ``$g$-theorem'' and
the operator contents.
We explained Grimm's results on the generalized defects,
applying our observations.

A possible extention of our present work is the analysis of the
boundary states in the general critical Ashkin-Teller model,
i.e. for general value of the compactification radius of the orbifold.
While two continuous families and eight discrete boundary
states exist in general, their relative stability will vary.
Moreover, for special values of the radius, additional boundary
states may appear, as in the case of the free boson~\cite{Callan:SU2boundary}.

In Ref.~\cite{WongAffleck}, the first attempt was made to
apply boundary conformal field theory to the defect problem
via folding.
However, it was essentially reproduction of the earlier results.
On the other hand, our present result provides a more detailed description
of defect lines in the Ising model, than that in the earlier literatures.
This demonstrates the power of boundary conformal field theory
for the application to the defect problems.
In particular, the solution indeed posesses $c=1$ structure
which is absent in the original $c=1/2$ problem before the folding.
Thus the folding seems a really necessary procedure to treat
the defect problem in terms of conformal field theory.
From the general viewpoint on boundary conformal field theory,
the present model seems one of the simplest nontrivial models with an infinite
number of boundary conformal towers.
Cardy's original treatment~\cite{Cardy:BCFT,Cardy:fusion}
mainly focused on minimal models which have
a finite number of conformal towers.
Most of the successful applications~\cite{Affleck:reviewKondo}
to $c \geq 1$ have also been to theories
with a finite number of primaries with respect to current algebra.
When the number of primaries cannot be reduced to a finite one,
there are several technical difficulties~\cite{WongAffleck}.
In spite of those difficulties, we found that the present problem is
tractable, though we have not yet completely understood the structure
of the theory as we do in minimal models.
We hope that the observations on the present model will be useful
in the further development of boundary conformal field theory
with infinitely many conformal towers.

\section*{Acknowledgments}

We thank F. Lesage for useful information.
This work is partly supported by NSERC of Canada.
M. O. thanks the
UBC Killam fellowship for financial support.

\appendix

\section*{Appendix: free energy and the boundary condition changing
operator}

\def\thesection{A}

By folding, a finite length defect line is mapped to a changed
boundary condition for a finite interval.
Let us discuss the free energy associated to such a finite interval
of changed boundary condition.
Using the transfer matrix (Hamiltonian) in the direction
parallel to the boundary (defect line),
the partition function of the system may be written as
\begin{equation}
  Z(l) = {\rm Tr} e^{-(L-l) H_0 } e^{ - l H_D },
\end{equation}
where $l$ is the interval of the changed boundary condition
(length of the defect), $L$ is the system size in the direction
parallel to the boundary and $H_0$ and $H_D$ are Hamiltonian
for the original and changed boundary conditions.
The partition function in the absence of the defect line is
\begin{equation}
  Z_0 = {\rm Tr} e^{-L H_0}  .
\end{equation}

On the other hand, a change in the boundary condition can be
associated with a ``boundary condition changing operator''~\cite{Cardy:fusion}.
The boundary condition changing operator acts in an enlarged Hilbert
space with all possible boundary conditions, and maps a whole subspace
with a particular boundary condition into each other.
By the boundary condition changing operator $U$,
the Hamiltonian is mapped as
\begin{equation}
  U^{\dagger} H_0 U = H_D
\end{equation}
In their pioneering work, Schotte and Schotte~\cite{Schotte2}
introduced such an operator for the free boson.
(See also~\cite{Affleck:bcchange}.)
Here we discuss the non-orbifold free boson.
The boundary condition changing operator that maps the
Dirichlet boundary condition $\varphi = 0$ to the Dirichlet
boundary condition with a different boundary value $\varphi = \delta$
is given as
\begin{equation}
  U = \exp{\left[ i \frac{\delta}{\pi} \tilde{\varphi}(0) \right]},
\label{eq:Udef}
\end{equation}
where $\tilde{\varphi}(0)$ is the quantum operator corresponding to the
$\tilde{\varphi}$ at the boundary.
Here the operators are quantized, regarding the direction parallel
to the boundary as a imaginary time.
As they claimed, $U$ is apparently a unitary operator.
Then it follows
\begin{equation}
  U^{\dagger} e^{- l H_0} U = e^{-l H_D},
\label{eq:Uexp}
\end{equation}
and thus
\begin{equation}
  Z(l) = {\rm Tr} \left[
             e^{-(L-l) H_0} U^{\dagger} e^{-l H_0} U \right] .
\end{equation}
Namely, the ratio of $Z(l)$ and $Z_0$ is identified with the two-point
correlation function of the boundary condition changing operator:
\begin{equation}
  \frac{Z(l)}{Z_0} = \langle U^{\dagger}(l)  U(0) \rangle ,
\label{eq:pftouu}
\end{equation}
where the expectation value is defined in the original theory
with the Hamiltonian $H_0$.
In the following, we assume the thermodynamic limit $L \rightarrow \infty$.
The asymptotic behavior for large $l$ is governed by
the lowest dimension $\Delta_b$
of the boundary condition changing operator.
This gives
\begin{equation}
 \langle U^{\dagger}(l)  U(0) \rangle  \sim \mbox{const.} l^{-\Delta_b}
\label{eq:Ucorr}
\end{equation}
The free energy of the changed boundary condition (defect line)
of a finite length is given by taking the logarithm as
\begin{equation}
  F(l) = - k T \log{\frac{Z(l)}{Z_0}} = \mbox{const.} + k T \Delta_b \log{l} .
\end{equation}
The constant term is non-universal.
From the dimensional analysis, we expect it to depend on the
short-distance cutoff (lattice spacing) $a$ as roughly $ - \log{a}$
In general, there can be also a non-universal shift in the
energy for a different boundary condition.
This contributes extra $A l$, where $A$ is a non-universal constant
of order $1/a$, to the free energy.

However, there is a subtlety in the above argument:
$U$ cannot be a unitary operator in the standard sense,
since $H_0$ and $H_D$ have {\em different} spectrum in general.
This peculiarity is presumably related to the presence of
infinite degrees of freedom in the field theory.
Thus the identity~(\ref{eq:Uexp}) and the above argument may
be questionable.
We argue that $U$ is still a unitary operator in some weak sense
and the above discussion on free energy is valid in general.
For the free boson case, we can demonstrate the validity
by calculating both sides of the equation~(\ref{eq:pftouu})
independently.

Let us consider the free boson with the Lagrangian~(\ref{eq:bosonLag}).
We take the boundary as $x$-axis ($y=0$) and assume that the system
lies in the upper half plane.
We consider the Dirichlet boundary condition $\varphi = \delta$
for the interval $0<x<l$, and $\varphi = 0$ elsewhere.
The corresponding boundary condition changing operator
is given in~(\ref{eq:Udef}).
The asymptotic form of the
two-point correlation function of the boundary condition changing
operator is given by~(\ref{eq:Ucorr}) with
$\Delta_b = \delta^2 / \pi^2 $.
On the other hand, the Free energy in the absence of the changed
boundary condition is given by
\begin{equation}
  Z_0 = \int {\cal D} \varphi
    e^{- \frac{1}{2 \pi} \int d^2 x (\partial_{\mu} \varphi)^2},
\label{eq:Z0}
\end{equation}
where $\varphi = 0$ at the boundary ($y=0$).
In the presence of the changed boundary condition,
we write
\begin{equation}
  \varphi = \varphi_c + \varphi_q,
\end{equation}
where $\varphi_c$ satisfies
\begin{eqnarray}
  \partial^2 \varphi_c &=& 0 \nonumber \\
  \varphi_c(x,0) &=& \left\{ \begin{array}{cl} 0 & (x < 0 , l<x) \\
                                             \delta & ( 0< x < l) 
                   \end{array}                                             
                   \right.
\\
  \varphi_c(x,y) & \rightarrow & 0 \;\;\; (y \rightarrow \infty)
\nonumber
\end{eqnarray}
$\varphi_q$ is only required to satisfy $\varphi_q =0$ at the boundary,
which is exactly the same condition $\varphi$ obeyed in the absence
of the changed boundary condition.
In other words, the change in the boundary condition is completely
absorbed by $\varphi_c$.
The partition function in the presence of the changed boundary condition
is given as
\begin{equation}
  Z(l) = \int {\cal D} \varphi_q 
  e^{ - \frac{1}{2 \pi} \int d^2x [ \partial_{\mu} (\varphi_q + \varphi_c)]^2 } .
\end{equation}
The action can be decoupled as
\begin{equation}
  \frac{1}{2 \pi} \int d^2 x (\partial_{\mu} \varphi_c)^2 +
                        (\partial_{\mu} \varphi_q )^2 ,
\end{equation}
since the cross term is shown to vanish by partial integration.
Actually there is an infinite number of solution for $\varphi_c$
and we must sum over such allowed $\varphi_c$,
apart from the integration over $\varphi_q$.
However, in the large-$l$ limit, only the one with minimum action
dominates.
For $|\delta| < \pi r$, $\varphi_c$ with minimum action is given by
\begin{equation}
  \varphi_c = \frac{\delta}{\pi}\left(
               \tan^{-1}{\frac{y}{x-l}} - \tan^{-1}{\frac{y}{x}} \right) .
\label{eq:phic}
\end{equation}
(The ones with larger value of the action is given by replacing $\delta$ by
$\delta + 2  \pi r$, etc.)
The leading term in the partition function is given by
\begin{eqnarray}
  Z(l) &=& \int {\cal D} \varphi_q 
            \exp{[ -  \frac{1}{2 \pi} \int d^2 x (\partial_{\mu} \varphi_c)^2 +
                     (\partial_{\mu} \varphi_q )^2 ]}
\nonumber \\
&=& \exp{ [ - \frac{1}{2 \pi} \int d^2 x (\partial_{\mu} \varphi_c )^2 ]}
  \int {\cal D} \varphi_q
  \exp{ [ - \frac{1}{2 \pi} \int d^2 x (\partial_{\mu} \varphi_q )^2 ]}
\end{eqnarray}
The functional integral over $\varphi_q$ is identical
to $Z_0$ in eq.~(\ref{eq:Z0}).
The action of $\varphi_c$ can be calculated using the explicit
form~(\ref{eq:phic}).
Actually it diverges at the singularities $(x,y)=(0,0)$ and $(l,0)$.
Introducing the short-distance cutoff $a$, we obtain
\begin{equation}
   \frac{1}{2 \pi} \int d^2 x (\partial_{\mu} \varphi_c )^2
       = \frac{ \delta^2}{\pi^2} \log{\frac{l}{a}} .
\end{equation}

Thus we finally get
\begin{equation}
  \frac{Z(l)}{Z_0} = (\frac{a}{l})^{\delta^2/\pi^2} ,
\end{equation}
which agrees with the two-point correlation function of the boundary
condition changing operator.
Thus the validity of the relation~(\ref{eq:pftouu}) is explicitly
shown in the free boson case.

This calculation can be generalized to the orbifold boson.
On the other hand, in general conformal field theories,
it is difficult to calculate directly the partition
function in the presence of the changed boundary condition.
We expect that the relation~(\ref{eq:pftouu}) also holds in
such general cases.

\end{document}